\documentclass[12pt]{article}

\usepackage{amsmath,amsfonts,amssymb,amstext}
\usepackage[overload]{empheq}

\usepackage[normal]{subfigure}
\usepackage{epsfig}
\usepackage{graphicx}
\usepackage{enumerate}
\usepackage{mathtools}
\usepackage{color}
\usepackage[normalem]{ulem}
\usepackage{float}
\usepackage{amsthm}

\usepackage{ulem}

\usepackage{hyperref}
\definecolor{purple}{rgb}{0.75,0,0.25}
\hypersetup{
  colorlinks   = true, 
  urlcolor     = blue, 
  linkcolor    = blue, 
  citecolor   = purple 
}

\usepackage[titletoc,title]{appendix}

\allowdisplaybreaks

\input epsf.tex

\newcommand{\wb}[1]{\overline{#1}}

\newcommand{\abs}[1]{\left\vert #1 \right\vert}

\newtheorem{remark}{Remark}

\def\R{\mathbb{R}}

\def\vu{\boldsymbol{u}}

\def\vn{\boldsymbol{n}}

\def\a{\alpha}

\def\r{\rho}

\def\T{\tau}

\def\p{\partial}

\def\bS{\boldsymbol{\sigma}}
\def\bT{\boldsymbol{\tau}}

\def\dint{\displaystyle\int}

\def\dfrac{\displaystyle\frac}

\def\bd{\boldsymbol}

\def\wtd{\widetilde}
\def\wht{\widehat}

\def\F{\mathcal{F}}

\definecolor{green3}{rgb}{0,0.8,0}
\title{A weakly non-hydrostatic shallow model for dry granular flows}
\author{J. Garres-D{\'i}az \thanks{Dpto. Matem{\'a}ticas. Edificio Einstein - Universidad de C\'ordoba.
  Campus de Rabanales, 14014-C\'ordoba, Spain, (jgarres@uco.es, tomas.morales@uco.es)}, E.D. Fern\'andez-Nieto \thanks{Dpto. Matem\'atica Aplicada I. ETS Arquitectura - Universidad de Sevilla.
  Avda. Reina Mercedes S/N, 41012-Sevilla, Spain, (edofer@us.es)}, A. Mangeney
\thanks{Institut de Physique du Globe de Paris, Seismology team, University Paris-Diderot, Sorbonne Paris Cit\'e, 75238, Paris, France, (mangeney@ipgp.fr)} \thanks{ANGE team, CEREMA, INRIA, Lab. J. Louis Lions, 75252, Paris, France}, T. Morales de Luna$^*$ }

\begin{document}
\date{}
\maketitle

\abstract{
A non-hydrostatic depth-averaged model for dry granular flows is proposed, taking into account vertical acceleration. A  variable friction coefficient based on the $\mu(I)$ rheology is considered. The model is obtained from an asymptotic analysis in a local reference system, where the non-hydrostatic contribution is supposed to be small compared to the hydrostatic one. The non-hydrostatic counterpart of the pressure may be written as the sum of two terms: one corresponding to the stress tensor and the other to the vertical acceleration. The model introduced here is weakly non-hydrostatic, in the sense that the non-hydrostatic contribution related to the stress tensor is not taken into account due to its complex implementation. The motivation is to propose simple models including non-hydrostatic effects.
In order to approximate the resulting model, a simple and efficient numerical scheme is proposed. It consists of a three-step splitting procedure, and it is based on a hydrostatic reconstruction, which allows us to obtain a well-balanced scheme. Two key points are: (i) the friction force has to be taken into account before solving the non-hydrostatic pressure. Otherwise, the incompressibility condition is not ensured; (ii) both the hydrostatic and the non-hydrostatic pressure are taken into account when dealing with the friction force. The model and numerical scheme are then validated based on several numerical tests, including laboratory experiments of granular collapse. The influence of non-hydrostatic terms and of the choice of the coordinate system (Cartesian or local) is analyzed. We show that non-hydrostatic models are less sensitive to the choice of the coordinate system. In addition, the non-hydrostatic Cartesian model produces deposits similar to the hydrostatic local model as suggested by Denlinger $\&$ Iverson \cite{denlinger:2004}, the flow dynamics being however different. Moreover, the proposed model, when written in Cartesian coordinates, can be seen as an improvement of their model, since the vertical velocity is computed and not estimated from the boundary conditions. In general, the non-hydrostatic model introduced here much better reproduces granular collapse experiments compared to hydrostatic models, especially at the beginning of the flow. The error on the thickness distribution for $\theta \leq 16^\circ$, at final time, is around 15$\%$ for the hydrostatic model, while it is approximately 7$\%$ for the non-hydrostatic model. An important result is that the simulated mass profiles up to the deposit and the front velocity are greatly improved. As expected, the influence of the non-hydrostatic pressure is shown to be larger for small values of the slope.
}

\bigskip



\frenchspacing   
\newpage
\section{Introduction}
Granular flows have been intensely studied in recent years, since they play an important role in the understanding of natural hazards (avalanches, submarine landslides,...) and industrial processes. Aerial granular flow models as well as other more complex models such as debris flows have been widely studied (e.g. \cite{jackson:2000,pitman:2005,iverson:2014,bouchut:2016}). The physical description of these type of flows is a very active field of research from several points of view. On the one hand, the definition of rheological laws describing the complex dynamics of the flow is a challenge nowadays, namely the solid-fluid transition occurring when a granular material is flowing. On the other hand, the mathematical modelling of these flows is a difficult issue, since the stress tensor has a complex expression usually, and therefore its numerical treatment is not straightforward. \\

From the physical point of view, the $\mu(I)$ rheology, introduced in Jop, Forterre $\&$ Pouliquen \cite{jop:2006,jop:2007}, is the most accepted rheological law describing dry granular flows. It considers a pressure and strain-rate dependent viscosity, through a variable friction coefficient depending on the inertial number. In addition, in recent years, some works have been devoted to improving this law by adding non-local effects to the $\mu(I)$ rheology. However, there are still many open questions around these non-local models (see e.g. \cite{pouliquen:2009,bouzid:2013}). Thus, the local $\mu(I)$ rheology continues being a very popular law for physicist when describing dry granular flows. \\

The $\mu(I)$-rheology was implemented in a 2D continuous model solving the full Navier-Stokes equations by Staron, Lagr\'ee $\&$ Popinet \cite{staron:2012} by using a regularization method to describe the static behavior of the material. By fitting the rheological parameters of the $\mu(I)$-rheology down to values smaller that those of the granular material involved, they were able to reproduce 2D discrete elements simulations. Ionescu {\it et al.} \cite{ionescu:2015} and Martin {\it et al.} \cite{martin:2017} quantitatively reproduced laboratory experiments of a granular collapse problem using finite element discretizations of the full 2D equations and an Augmented Lagrangian method, as well as a simplified description of the lateral wall effects. Bouchut {\it et al.} \cite{bouchut:2016a} derived an analytic expression for the non-hydrostatic pressure. It is based on an asymptotic analysis under some hypothesis, such as shallow flow and small velocity. As a consequence, only the terms related to the stress tensor are considered in the definition of the non-hydrostatic pressure counterpart, while the acceleration in the direction normal to the slope is neglected. Comparison of this analytical formulation with the pressure, computed by a model solving the full 2D Navier-Stokes equation, showed that these non-hydrostatic analytical terms describe well part of the non-hydrostatic pressure (see Figure 18 of \cite{martin:2017}). \\

It is a well-known fact that the computational cost of solving flows with a moving free surface with a 3D (or 2D)  solver is huge.  Shallow depth-averaged and hydrostatic layer-averaged models have been widely used in order to reduce this computational effort. These models are mainly based on the pioneering work of Savage-Hutter \cite{savage:1989}, where the friction between the bottom and the granular material is modelled though a constant Coulomb friction coefficient. Pouliquen \cite{pouliquen:1999a,pouliquen:1999b} proposed to replace this constant value by a friction coefficient depending on the strain rate. In a more recent work, Pouliquen $\&$ Forterre \cite{pouliquen:2002} proposed to make this coefficient depend on the Froude number. The resulting model was used by Mangeney-Castelnau {\it et al.} \cite{mangeney:2003} and Mangeney {\it et al.} \cite{mangeney:2007} to simulate granular flows on simple topography, making it possible to reproduce qualitatively self-channeling flows and levee formation. It has also been successfully used to simulate real landslides over complex topography (e. g. \cite{pirulli:2008}).
More recently, Gray $\&$ Edwards \cite{gray:2014} proposed a slightly modified depth-averaged model by including second-order viscous terms derived by assuming a Bagnold profile for the downslope velocity. It reduced to the model of \cite{pouliquen:1999a} when these second-order terms are dropped. This model, combined with the friction coefficient \cite{pouliquen:2002}, is used in Edwards $\&$ Gray \cite{edwards:2015} to simulate roll-waves and erosion-deposition waves and in Baker {\it et al.} \cite{Baker:2016}, making it possible to recover the transversal profile of the downslope velocity. However, the main drawback of this model is the fact that the velocity profile is prescribed even though the shape of the velocity profile is known to change during the flow. This change can be handled using multilayer models as done by Fern\'andez-Nieto {\it et al.} \cite{fernandezNieto:2016,fernandezNieto:2018} that also used the $\mu(I)$ rheology. Indeed such models have been shown to reproduce the observed change in velocity profiles during granular flows on inclined planes. Using multilayer models, \cite{fernandezNieto:2016} also showed that the $\mu(I)$ rheology better reproduces the dynamics of granular flows than using a constant friction coefficient.\\

In all these depth-averaged or multilayer models, the pressure is assumed to be hydrostatic. In addition, as explained before, the analytic formula for the pressure proposed by \cite{bouchut:2016a} includes the rheology terms but not the normal acceleration terms due to their assumption of small velocity flows. Furthermore, it is well known that the initial dynamics of granular collapse is not well reproduced by shallow depth-averaged models, in particular because of the importance of non-hydrostatic effects in this regime (e. g. \cite{mangeney:2005}, \cite{fernandezNieto:2016,fernandezNieto:2018}). Therefore, a non-hydrostatic shallow model for granular flows, which takes into account the acceleration in the direction normal to the slope may significantly improve the ability of depth-averaged models to reproduce flow regimes where non-hydrostatic effects are important such as during the first instant of granular collapses.\\

Non-hydrostatic shallow water models have been a popular topic of research in recent years. The idea is to  improve nonlinear dispersive properties of water waves by including some information on the vertical structure of the model. One way of doing so is by including a non-hydrostatic pressure in the model. In the usual process of averaging the fully 3D equations, the pressure is no longer assumed to be hydrostatic and is split into a hydrostatic and a non-hydrostatic part (see for instance \cite{casulli:1999}, \cite{stelling:2003}, \cite{sainteMarie:2015}, and \cite{yamazaki:2008}, among others). The advantage of non-hydrostatic models when compared to classical dispersive systems is that they present only first-order derivatives, which are easier to treat numerically (see e.g. \cite{escalante:2020}). Moreover, the particular structure of these type of models and
their similarities with shallow water equations allow extending many well-known numerical schemes for shallow water equations to non-hydrostatic models, see for instance \cite{escalante:2018}. In view of the improvement and possibilities of this technique for shallow waters, one could think that a similar approach would be interesting for granular flows.\\

The choice of the coordinate system plays a key role, in particular in depth-averaged models, which are obtained after an integration procedure. If a Cartesian coordinate system is chosen, the $3$D model is integrated along the vertical Cartesian direction. However, it is usual in geophysical flows to use local coordinates (see e.g. \cite{pirulli:2008,gray:2014,bouchut:2016}), where the integration is made along the normal direction to the topography, typically a reference plane with constant slope (although it may vary along the domain \cite{fernandezNieto:2008}) or an arbitrary topography \cite{bouchut:2004b,mangeney:2007}. These models are more accurate for granular flows since the computed velocity is tangent to the topography, which is physically relevant, in contrast to Cartesian models. Recently, Delgado-S\'anchez {\it et al.} \cite{delgadoSanchez:2020} proposed a two-layer depth-averaged model, where they use Cartesian coordinates for an upper water layer and local coordinates for a lower granular layer since for water waves the vertical acceleration can be supposed to be small while for granular flows the acceleration normal to the slope is small. They showed that large errors are obtained when the coordinate system is not correctly chosen. Denlinger $\&$ Iverson \cite{denlinger:2004} proposed a Cartesian model for landslides, where the pressure is corrected by an approximation of the vertical acceleration. They show that the results of this Cartesian model are close to the results of a hydrostatic model in local coordinates for dam break analytical solutions. An approximation of the vertical acceleration is introduced to approximate the non-hydrostatic pressure in  \cite{denlinger:2004}, by taking the average between the vertical velocity derived from the free surface and bottom boundary conditions. Nevertheless, the vertical velocity is not computed as an unknown of the system.

\medskip

In this paper we deduce a simple model for granular flows including non-hydrostatic effects related to the acceleration in the direction normal to the slope. Then, a simple and efficient numerical scheme will be proposed that will allow us to notably improve the results of hydrostatic models. To our knowledge, this is the first non-hydrostatic shallow model (computing the normal acceleration) for dry granular flows.
It follows from an asymptotic analysis and the decomposition of the pressure into a hydrostatic pressure and a small perturbation (non-hydrostatic contribution). For simplicity, the non-hydrostatic pressure will be assumed to follow a linear profile in the normal direction. Note that this is a simplification so that the final system is easier to deal with. Nevertheless, other type of profiles could be used, which would result in a more complex model, including extra variable and equations (see for instance the approaches used in \cite{fernandezNieto:2018a,escalante:2020}).  The model will take into account a bottom friction coefficient defined by the $\mu(I)$ rheology.
Although the model will be derived using local coordinates, one may follow easily the same procedure in order to obtain a similar version in Cartesian coordinates. In \cite{denlinger:2004}, a Cartesian non-hydrostatic model is also proposed. The main differences are: (i) in the proposed model  the vertical velocity is an unknown, whereas in \cite{denlinger:2004} it is estimated in terms of the kinematic boundary conditions; (ii)  the non-hydrostatic pressure correction in the proposed model is the Lagrange multiplier associated to averaged incompressibility equation, while in \cite{denlinger:2004} it is approximated from the total time derivative of the estimated vertical velocity.

\medskip 

This paper is organised as follows. In Section \ref{se:derivation} we present the initial system and the derivation of the non-hydrostatic model, based on an asymptotic analysis. Section \ref{se:numericalapproximation} is devoted to the development of an efficient numerical scheme to approximate the proposed non-hydrostatic model. This is done with a three-step splitting technique, where the friction term is applied before solving the non-hydrostatic pressure. This is one of the key points of the scheme. In Section \ref{se:tests} different numerical tests are presented. The objective is to show the influence of the choice of the coordinate system (Cartesian or local) for hydrostatic and non-hydrostatic models. We also present a comparison with experimental data of granular collapses over inclined planes, showing that the non-hydrostatic model gives better results than the hydrostatic one, especially at short times and more generally on the mass profiles up to the deposit. In addition, non-hydrostatic models make it possible to include the vertical velocity as a variable in the model and thus to simulate the effect of the opening gate in the laboratory experiments, which is not possible with hydrostatic model. Finally, some conclusions are presented in section \ref{se:conclusions}.

\section{Derivation of a non-hydrostatic shallow $\mu(I)$-model}\label{se:derivation}
In this section we deduce the non-hydrostatic model. It follows from an
asymptotic analysis of the 2D Navier-Stokes system and the integration of the
resulting equations along the normal direction to the topography.

\subsection{Initial system}
\begin{figure}[!ht]
	\begin{center}
		\includegraphics[width=0.51\textwidth]{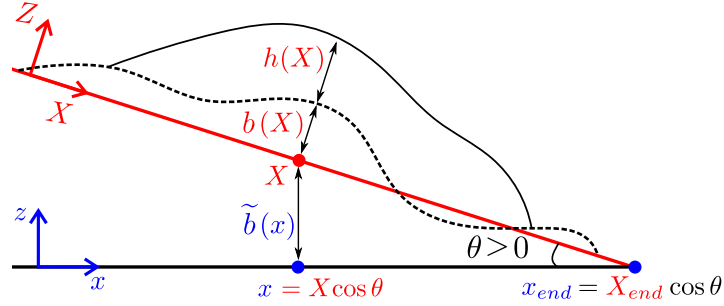}
	\end{center}
	\caption{\label{fig:local_esq} \it{Sketch of the Cartesian (blue) and local (red) reference system.}}
\end{figure}

First, let us establish the notation used in this paper. In particular, we
shall consider two different reference systems. We shall use local (or tilted) coordinates, as is usually done  in  granular flows, as well as Cartesian
coordinates. We shall denote by $(x,z)$ the Cartesian coordinate system, while
$(X,Z)$ will denote the local coordinates. This local coordinates will refer
to a given fixed inclined plane. More explicitly, let us consider a reference
inclined plane with slope $\theta$, that is a plane given by the function
$\wtd{b}(x) = (x_{end}-x) \tan\theta$, where $x_{end}$ is the ending point of the
domain. Let us remark that we consider here the usual convention in geophysical applications
which establishes that a positive angle $\theta$ corresponds to a negative
slope. Local coordinates $(X,Z)$ are then considered, measured along the
downslope and normal direction to the reference plane, $\wtd{b}(x)$,
respectively. The velocity vector is $\vu=(u,w)$, where $u$ is the downslope
component of the velocity and $w$ is the normal one. Finally, we consider also
a bottom topography $b(x)$ over the reference inclined plane (see Figure \ref{fig:local_esq}).

As starting point, we consider the 2D Navier-Stokes system for a flow with constant density $\rho$ given by
\begin{equation}\label{eq:NS_2D}
\left\{
\begin{array}{l}
\displaystyle \p_{X} u+ \p_{Z}w= 0,\\[3mm]
\displaystyle \p_{t}u+u\,\p_{X}u + w\,\p_{Z}u+\dfrac{1}{\r}\p_{X}(p_T) =   g\sin\theta+\dfrac{1}{\r}\big(\p_{X}(\T_{XX})+\p_{Z}( \T_{XZ})\big),\\[3mm]
\displaystyle \p_{t}w+ u\,\p_{X}w + w\,\p_{Z}w +\dfrac{1}{\r}\p_{Z}(p_T)=- g\cos\theta+\dfrac{1}{\r}\big(\p_{X} (\T_{ZX}) +\p_{Z}(\T_{ZZ})\big),
\end{array}
\right.
\end{equation}
where $g$ is the gravity force, $p_T$ is the total pressure and
$$\bT = \left(\begin{matrix}
\T_{XX} & \T_{XZ}\\
\T_{ZX} & \T_{ZZ}
\end{matrix}\right)$$
is the deviatoric part of the total stress tensor, $\bS = -p_T {\bd I}+\bT$, where $\bd I$ is the identity matrix.

The total pressure shall be decomposed as sum of the hydrostatic and the non-hydrostatic contribution (see e.g. Casulli \cite{casulli:1999})
\begin{equation}
\label{eq:totalpressure}
p_T = p_H + p_{NH},
\end{equation}
where $p_H=g\cos\theta\left(b+h-z\right)$ is the hydrostatic pressure and $p_{NH}$ denotes the non-hydrostatic counterpart. The atmospheric pressure has been set to zero for the sake of simplicity and we shall assume that the non-hydrostatic pressure vanishes at the surface
$$p_{NH|_{b+h}} = p_{H|_{b+h}} = p_{T|_{b+h}} = 0.$$
We also assume that  ${\T_{XZ} }_{|_{b+h}} = 0.$
We consider at the free surface the usual kinematic condition
\begin{equation}\label{eq:BC_kinematicCond}
\p_t h + u_{|_{b+h}}\p_X b - w_{|_{b+h}} = 0.
\end{equation}
At the bottom, we have the non-penetration condition
\begin{equation}\label{eq:BC_nopen}
\vu\cdot {\bd n}^b = 0,
\end{equation}
where ${\bd n}^b = \left(\p_{X}b,-1\right)$ is the downward normal vector to the bottom.

In addition, a Coulomb-type friction condition is considered:
\begin{equation}\label{eq:BC_bottom}
\bS\ \vn^{b} - \left(\left(\bS\ \vn^{b}\right)\cdot\vn^{b}\right)\vn^{b} = \left(\begin{matrix}
-\mu \,p_T\dfrac{u}{\abs{u}}, \ 0\
\end{matrix}\right)',
\end{equation} where here $\mu$ denotes the friction coefficient This coefficient may be constant (i.e., Savage-Hutter model \cite{savage:1989}) or variable according to some other rheological laws. Currently, the $\mu(I)$ rheology (see e.g. \cite{jop:2006}) is the most accepted law describing dry granular flows. Therefore, we shall consider this rheology and define
\begin{equation}\label{eq:muI}
\mu = \mu(I) = \mu_s + \dfrac{\mu_2-\mu_s}{I_0 + I}\,I,
\end{equation}
where $\mu_s,\mu_2,I_0$ are constant values, and $I$ is the inertial number defined as
$$
I = \dfrac{2d_s\|D(\vu)\|}{\sqrt{p_T/\r_s}}.
$$
In the previous equation, $d_s$ is the particle diameter, $\r_s$ the particle density, and $D(\vu)$ is the strain-rate tensor with $\|D(\vu)\| = \sqrt{0.5 D:D}$. Note that the apparent flow density is $\r = \r_s \varphi_s$, where $\varphi_s$ is the solid volume fraction.
This rheological law is included in system \eqref{eq:NS_2D} by defining the deviatoric stress tensor $\bT = \nu D(\vu)$, where the   viscosity coefficient, $\nu$, is defined according to the $\mu(I)$ rheology as (see e.g. \cite{lagree:2011,fernandezNieto:2016})
\begin{equation}
\label{eq:viscosity_muI}
\nu = \dfrac{\mu(I)p_T}{\|D(\vu)\|}.
\end{equation}

\subsection{Dimensional analysis and derivation of the model}
We follow the typical dimensional analysis for dry granular flows (see e.g. \cite{gray:2014,fernandezNieto:2018}) to obtain a simplified shallow model. Therefore, the ratio between the characteristic height $(H)$ and length $(L)$ is assumed to be small
$$
\varepsilon = \dfrac{H}{L}.
$$
We denote as well by $U$ the characteristic velocity. In what follows, we will denote with tildes ($\wtd{\cdot}$) the non-dimensional variables. Then, we have:
\begin{equation*}
\label{nondim_var}
\begin{array}{c}(X,Z,t) = (L\wtd{X},H\wtd{Z},(L/U)\wtd{t}),\\ \\
(u,w) =  (U\wtd{u}, \varepsilon U\wtd{w}),\\
\\
h = H\wtd{h},
\quad
\r = \r_{0}\wtd{\r},\quad
p_T = \r_{0}U^{2}\wtd{p_T},\\
\\
\left(\T_{XX},\T_{XZ},\T_{ZZ}\right) = \r_{0}U^2\left(\varepsilon\wtd{\T_{XX}},\wtd{\T_{XZ}},\varepsilon\wtd{\T_{ZZ}}\right).
\end{array}
\end{equation*}

Note also that
$$
D(\vu) = \dfrac{U}{H}\ \frac12 \left( \begin{matrix}
2\varepsilon \p_{\wtd{X}} \wtd{u} &          \p_{\wtd{Z}}\wtd{u}+\varepsilon^2\p_{\wtd{X}}\wtd{w}\\
\\
\p_{\wtd{Z}}\wtd{u}+\varepsilon^2\p_{\wtd{X}}\wtd{w} &   2\varepsilon\p_{\wtd{Z}}\wtd{w}
\end{matrix}\right),\quad \mbox{and}\quad \|D(\vu)\| = \abs{\p_Z u}/2
$$
up to first order. Defining now the Froude number as $Fr = U/\sqrt{gH\cos\theta }$ and dropping tildes for the sake of simplicity, system \eqref{eq:NS_2D} is written in non-dimensional form as
\begin{subequations}\label{eq:nondim_NS_3D}
	\begin{align}[left = \empheqlbrace\,]
	&\p_{X} u+\p_{Z}w= 0,\label{eq:nondim_NS_3Da}\\[3mm]
	&\r\big(\p_{t}u+u\p_{X} u +  w\p_{Z}u\big)+\p_{X}(p_T)  =  \dfrac{1}{\varepsilon} \dfrac{\r}{Fr^2}\tan\;\theta+\varepsilon\p_{X}(\T_{XX})+ \dfrac{1}{\varepsilon}\p_{Z}(\T_{XZ}), \label{eq:nondim_NS_3Db}\\[3mm]
	&\r\varepsilon^2\big(\p_{t}w+ u\,\p_{X}w + w\,\p_{Z}w\big) +\p_{Z}(p_T)=- \dfrac{\r}{Fr^2}+\varepsilon\p_{X} (\T_{ZX})  +\varepsilon\p_{Z}(\T_{ZZ}).\label{eq:nondim_NS_3Dc}
	\end{align}
\end{subequations}
In addition, the friction condition \eqref{eq:BC_bottom} at the bottom is given by
\begin{equation}\label{eq:BC_bottom_nodim}
\left(\begin{matrix}
\T_{XZ},\ 0
\end{matrix}\right)' = \left(\begin{matrix}
\mu(I) p_{T}\dfrac{u}{\abs{u}}, \ 0\
\end{matrix}\right)', \qquad \mbox{at}\ z=b.
\end{equation}

Finally, we assume that the non-hydrostatic pressure is smaller than the hydrostatic one. To this aim, we consider that the pressure takes the form
$$
p_T = p_H + \varepsilon q_1 + \varepsilon^2 q = \dfrac{\r}{Fr^2}\left(b+h-z\right) + \varepsilon q_1 + \varepsilon^2 q ,
$$
where $q_1,q$ are the first and second order terms of the non-hydrostatic counterpart. It leads to the vertical momentum conservation equation
\begin{equation}
\r\varepsilon^2\big(\p_{t}w+ u\,\p_{X}w + w\,\p_{Z}w\big) +\varepsilon\p_{Z}q_1+\varepsilon^2\p_{Z}q= \varepsilon\p_{X}( \T_{ZX} ) +\varepsilon\p_{Z}(\T_{ZZ}),
\end{equation}
where the gravitational term has been cancelled with the hydrostatic contribution of the pressure. Now, by comparing the terms with the same order of magnitude in previous equation, we obtain that
\begin{equation}\label{eq:pnh_antesdeintegrar_q1}
\p_Zq_1 = \p_{X} (\T_{ZX})  + \p_{Z}(\T_{ZZ} ),
\end{equation}
and
\begin{equation}\label{eq:pnh_antesdeintegrar}
-\p_{Z}q = \r\big(\p_{t}w+ u\,\p_{X}w + w\,\p_{Z}w\big).
\end{equation}
In this work, the aim is to obtain the simplest depth-averaged non-hydrostatic model, improving the results of hydrostatic models. First, from the numerical point of view, it is difficult to deal with the viscous terms in \eqref{eq:pnh_antesdeintegrar_q1}. Second, the problem of considering the vertical acceleration has been widely studied for shallow water flows, both from the modeling and the numerical point of view. Thus, we are going to neglect the first order terms of the pressure and keep the second order contribution, i.e., we keep \eqref{eq:pnh_antesdeintegrar} and do not consider \eqref{eq:pnh_antesdeintegrar_q1}. In the numerical tests, we will show that this choice significantly improves the results compared to the hydrostatic assumption.
\\\\

Next, we derive the final model by integrating \eqref{eq:nondim_NS_3Da},\eqref{eq:nondim_NS_3Db} and \eqref{eq:pnh_antesdeintegrar} along the vertical direction. To this aim, for any variable $f$ we define its average on the normal direction by
$$
\wb{f} = \dfrac1h \dint_{b}^{b+h} f dZ.
$$
We shall use that $\wb{f\cdot g} = \wb{f}\cdot\wb{g}$, which is true up to first order. Then, by integrating on the normal direction equations \eqref{eq:nondim_NS_3Da} and \eqref{eq:pnh_antesdeintegrar} between $b$ and $b+h$, and taking into account the Leibniz's rule, we get
$$\begin{array}{c}
\p_X\left(h\wb{u}\right) + \left(u_{|_{b}}\p_X b - w_{|_{b}}\right) - \left(u_{|_{b+h}}\p_X\left(b+h\right) - w_{|_{b+h}}\right) = 0,\\[4mm]
\p_t\left(h\wb{w}\right) + \p_X\left(h\wb{u}\,\wb{w}\right) + w_{|_{b}} \left(u_{|_{b}}\p_X b - w_{|_{b}}\right) - w_{|_{b+h}}\left(\p_t\left(b+h\right)+u_{|_{b+h}}\p_X\left(b+h\right) - w_{|_{b+h}}\right) = \dfrac{1}{\r} q_{|_{b}}.\end{array}
$$
Using now the kinematic and non-penetration conditions we get
\begin{equation}
\label{eq:eq_masa_nodimens}
\begin{array}{c}
\p_t h + \p_X\left(h\wb{u}\right)  = 0,\\[4mm]
\p_t\left(h\wb{w}\right) + \p_X\left(h\wb{u}\,\wb{w}\right) = \dfrac{1}{\r} q_{|_{b}}.\end{array}
\end{equation}
A closure relation is needed for the non-hydrostatic pressure. For the sake of simplicity, we shall assume that   $p(X,\cdot,t)$ has a linear profile. This hypothesis implies that $q_{|_{b}} = 2 \wb{q}$. Then, as a consequence, the system has only one extra unknown, $\wb{q}$, which is the Lagrange multiplier associated to the averaged incompressibility condition deduced  bellow.  Nevertheless, other possible choices may be made on the profile of the non-hydrostatic pressure. This would mean then that the system will have extra unknowns and equations (see e.g. \cite{fernandezNieto:2018a,escalante:2020}).

We focus now on the horizontal momentum equation \eqref{eq:nondim_NS_3Db} up to first order. Noticing that
$$
\dint_{b}^{b+h} \p_X p_T \,dz= \dfrac{\r}{Fr^2} h \p_X\left(b+h\right) + \varepsilon^2\left( \p_X\left(h\wb{q}\right) + q_{|_{b}}\p_X b  \right),
$$
and using $q_b = 2\wb{q}$, the depth-averaged momentum conservation equation is
\begin{equation}
\label{eq:ec_mom_hor_nodim}
\r\left( \p_t\left(h\wb{u}\right) + \p_X\left(h\wb{u}^2\right) + \dfrac{1}{Fr^2}h \p_X\left(h+b+\dfrac{x\tan\theta}{\varepsilon}\right) \right) = -\varepsilon^2\big(\p_X\left(h\wb{q}\right) + 2\wb{q}\p_X b\big) -\dfrac{1}{\varepsilon}\mu(I_{|_{b}}) p_{T|_{b}} \dfrac{\wb{u}}{\abs{\wb{u}}}
\end{equation}
where the friction condition \eqref{eq:BC_bottom_nodim} has been used.

Considering equations \eqref{eq:eq_masa_nodimens} and \eqref{eq:ec_mom_hor_nodim}, we have a system with $3$ equations and $4$ unknowns ($h,u,w,q$). Then, in order to close the system, we integrate the continuity equation \eqref{eq:nondim_NS_3Da} between the bottom ($b$) and the midpoint of the layer ($b+h/2$), obtaining (by using the Leibniz's rule and the non-penetration condition)
$$
\p_X \left( \dint_{b}^{b+h/2} u dz\right) - u_{|_{b+h/2}}\p_X\left(b+h/2\right) + w_{|_{b+h/2}} = 0.
$$
Notice that
$$\wb{u} = u_{|_{b+h/2}} + \mathcal{O}(\varepsilon^2),  \quad \dint_{b}^{b+h/2} u dz = h\wb{u} + \mathcal{O}(\varepsilon) \quad \mbox{and}\quad \wb{w} = w_{|_{b+h/2}} + \mathcal{O}(\varepsilon^2),$$
thanks to the midpoint and the rectangular quadrature rule to approximate those integrals. Then,
we obtain that
$$
h\wb{w} = h\wb{u}\p_X b - \dfrac{h^2}{2}\p_X \wb{u}.
$$

\subsection{Final model}
Collecting the equations that we have obtained in previous subsections and going back to dimensional variables, we get the system
\begin{subequations}\label{eq:finalmodel}
	\begin{align}[left = \empheqlbrace\,]
	&\p_t h + \p_X\left(h\wb{u}\right)  = 0,\label{eq:finalmodel_a}\\[3mm]
	&\r\left( \p_t\left(h\wb{u}\right) + \p_X\left(h\wb{u}^2\right) + g\cos\theta h \p_X\left(h+b+\wtd{b}\right) \right) = -\big(\p_X\left(h\wb{q}\right) + 2\wb{q}\p_X b\big) - \mu(I_{|_{b}}) p_{T|_{b}} \dfrac{\wb{u}}{\abs{\wb{u}}},\label{eq:finalmodel_b}\\[3mm]
	&\r\big(\p_t\left(h\wb{w}\right) + \p_X\left(h\wb{uw}\right)\big) =  2\wb{q},\label{eq:finalmodel_c}\\[3mm]
	&h\wb{w} = h\wb{u}\p_X b - \dfrac{h^2}{2}\p_X \wb{u}.\label{eq:finalmodel_d}
	\end{align}
	where $\mu(I)$ is given by \eqref{eq:muI}, and
	\begin{equation}\label{eq:finalmodel_e}
	I_{|_{b}} = \dfrac{d_s\abs{(\p_z u)_{|_b}}}{\sqrt{p_{T_{|_b}}/\r_s}},\quad\mbox{with}\quad p_{T|_{b}}/\r_s = \dfrac{\r}{\r_s} \left(g\cos\theta h + 2\dfrac{\wb{q}}{\r}\right) = \varphi_s \left(g\cos\theta h + 2\dfrac{\wb{q}}{\r}\right).
	\end{equation}
\end{subequations}

Note that the friction term is computed taking into account the total pressure, hydrostatic and non-hydrostatic. In order to simplify system \eqref{eq:finalmodel}, in what follows we shall redefine the non-hydrostatic variable as $\wb{q} = \wb{q}/\r$. In the next section, we detail the numerical scheme proposed to approximate system \eqref{eq:finalmodel}.

\begin{remark}\label{re:remark_modelo_hidrostatico}
	A hydrostatic version of model \eqref{eq:finalmodel} is obtained from \eqref{eq:finalmodel_a}, \eqref{eq:finalmodel_b} and \eqref{eq:finalmodel_e}, by setting $\bar{q}=0$. The resulting hydrostatic model corresponds to the one proposed in \cite{pouliquen:1999b}, which is the Savage-Hutter model \cite{savage:1989}, where the friction coefficient is improved by using the $\mu(I)$ rheology. This model also matches to the one proposed in \cite{gray:2014} when the the viscous term $\p_X \T_{XX}$ is removed.
\end{remark}

\begin{remark}
	Concerning steady states for system \eqref{eq:finalmodel}, we shall focus on stationary solution when the granular flow is at rest. More explicitly,
	\begin{equation}\label{eq:steadysolution}
	\bar{u} = \bar{w} = \bar{q} = 0, \quad \mbox{and}\quad \abs{\p_X\left(h+b+\wtd{b}\right)} \leq \mu_s.
	\end{equation}
	Note that previous equation corresponds to solutions at rest for classical Savage-Hutter model.
	When designing a numerical scheme for model \eqref{eq:finalmodel} we will be interested in preserving these steady states, that is, a well-balanced scheme for \eqref{eq:steadysolution}.
	
\end{remark}

\section{Numerical approximation}\label{se:numericalapproximation}
One of the aims of this paper is to propose a simple and efficient numerical scheme to approximate the previously introduced non-hydrostatic shallow $\mu(I)$-model \eqref{eq:finalmodel}. We propose a numerical approximation consisting in a three-steps method, where the main novelty is how to deal with the Coulomb friction term together with the non-hydrostatic pressure. The first step involves the hyperbolic part of the system and an explicit discretization of the non-hydrostatic term. In this first step  a path-conservative finite volume scheme is considered, together with a hydrostatic reconstruction in order to ensure the well-balance property. Secondly, the Coulomb friction is added taking into account also the non-hydrostatic contributions. Finally, a  non-hydrostatic pressure deviation in time is computed and the velocity field is corrected accordingly.

Regarding the computational cost of this non-hydrostatic model, it was shown that using the strategy in Escalante {\it et al.} \cite{escalante:2018}, the computational effort associated to this non-hydrostatic model is approximately $2.4$ times greater than the one for the hydrostatic version of the model. Similar results are expected here.
In what follows we shall describe each step in detail.

Let us denote by $\bd{w} = \left(h, hu,hw\right)^{'}$. We consider a usual Finite Volume discretization, where the horizontal domain is divided in control volumes $V_i = \left[x_{i-1/2},x_{i+1/2}\right]$, for $i\in\mathcal{I}$. For the sake of simplicity we assume a fixed volume mesh size $\Delta x$. We denote the center of each volume cell by $x_i = \left(x_{i-1/2}+x_{i+1/2}\right)/2$. For any time $t$, we consider the cell averages
$$
\bd{W}_i(t) = \dfrac{1}{\Delta x}\dint^{x_{i+1/2}}_{x_{i-1/2}} \bd{W}(x,t) dx.
$$
Regarding non-hydrostatic terms, a staggered grid is considered formed by the points $x_{i-1/2},x_{i+1/2}$ of the interfaces for each cell $V_i$. Let us denote the point values of the function $q$ representing the non-hydrostatic pressure on point $x_{i+1/2}$ at time $t$ by
$$q_{i+1/2}(t) = q(x_{i+1/2},t).$$
Remark that this corresponds to a second order approximation of the cell average of the pressure on the staggered grid $[x_i,x_{i+1}]$. In what follows and for the sake of simplicity, we omit the dependence on the time $t$.

In order to define a numerical scheme for \eqref{eq:finalmodel_hip}, we have to consider the following three key points:
\begin{itemize}
	\item The resulting scheme should be well-balanced for \eqref{eq:steadysolution}. This is achieved by means of a hydrostatic reconstruction procedure, taking into account friction terms (see \cite{fernandezNieto:2018}).
	\item The friction contribution should be taken into account before solving the non-hydrostatic pressure. Otherwise, the incompressibility condition is not ensured.
	\item Both, hydrostatic and non-hydrostatic pressures, should be considered when dealing with the friction term.
\end{itemize}
We propose here a numerical scheme based on three steps which are described in what follows.

\subsection*{Step 1: Hyperbolic problem}
The first step focuses on solving the hyperbolic system obtained when friction and non-hydrostatic effects are removed from system \eqref{eq:finalmodel}. Therefore, we obtain the following system (bars are dropped for simplicity):
\begin{subequations}\label{eq:finalmodel_hip}
	\begin{align}[left = \empheqlbrace\,]
	&\p_t h + \p_X\left(hu\right)  = 0,\label{eq:finalmodel_hip_a}\\[3mm]
	& \p_t\left(hu\right) + \p_X\left(hu^2\right) + g\cos\theta h \p_X\left(h+b+\wtd{b}\right) = 0,\label{eq:finalmodel_hip_b}\\[3mm]
	&\p_t\left(hw\right) + \p_X\left(huw\right) =  0.\label{eq:finalmodel_hip_c}
	\end{align}
\end{subequations}
We see that equations \eqref{eq:finalmodel_hip_a} and \eqref{eq:finalmodel_hip_b} correspond to a shallow water system, combined with a transport equation for a passive scalar \eqref{eq:finalmodel_hip_c}. In order to solve system \eqref{eq:finalmodel_hip}, we follow a similar approach as in \cite{fernandezNieto:2008b}. In particular, we follow the path-conservative framework \cite{pares:2006} to define a HLL-type method for the shallow water system in a similar way as it is done in \cite{fernandezNieto:2018}.
After that, the third equation is considered as a transport equation of a passive scalar.
For the sake of completeness, let us describe this in detail.

System \eqref{eq:finalmodel_hip_a}-\eqref{eq:finalmodel_hip_b} may be written in compact form as
\begin{align}
\label{eq:compact_form}
\p_{t}\bd{W} + \p_{X}\bd{F_c}(\bd{W}) +  \bd{S}(\bd{W})\p_{X} \left(\wtd{b} + b+h\right) = \bd{0}
\end{align}
where $\bd{W} = \left(h, hu,hw\right)^{'} \in \R^{3}$ is the unknown vector, $\bd{F_c}(\bd{W}) = \left(hu, hu^2,h u w\right)^{'}$ are the convective terms and $\bd{S}(\bd{W}) = \left(0, g\cos\theta h,0 \right)^{'}$  defines the source term which accounts for the hydrostatic pressure.

Now, for every interface $x_{i+1/2}$ we perform a hydrostatic reconstruction on
the variables $h$ and $Z$ (see \cite{audusse:2004}) by setting, to the left and
right of the interface respectively,
\begin{equation}\label{eq:recfondo}
\begin{array}{c}
h_{i+1/2}^- = \mbox{max}(0,h_i - (\Delta Z_{i+1/2})_+);\\[4mm]
h_{i+1/2}^+ = \mbox{max}(0,h_{i+1} - (-\Delta Z_{i+1/2})_+),
\end{array}\quad\mbox{with}\quad (\Delta Z_{i+1/2})_+ =
\mbox{max}(0,z_{b,i+1}-z_{b,i}).
\end{equation}

Then, the finite volume method is described as
\begin{equation} \label{eq:step1}
\begin{array}{l}
\bd{W}_{i}^{n+1/3}=\bd{W}_i^n +\dfrac{\Delta t}{\Delta x} \left( \bd{\F}_{i-1/2}^n-\bd{\F}_{i+1/2}^n \,
+ \dfrac12\left(\bd{\mathcal{S}}_{i+1/2}^n+\bd{\mathcal{S}}_{i-1/2}^n \right)\right),
\end{array}
\end{equation}
with
\begin{equation*}
\bd{\mathcal{S}}_{i+1/2}^n =  \dfrac{1}{2}(\bd{S}(\bd{W}_{i+1}^n)+\bd{S}(\bd{W}_{i}^n)) \bigl(h_{i+1/2}^{+,n} - h_{i+1/2}^{-,n}  \bigr),
\end{equation*}
where $h_{i+1/2}^\pm$ are the reconstructed heights.

Finally, the numerical flux corresponding to the convective terms $\bd{\F}_{i+1/2}^n$, is
\begin{equation*}
\begin{array}{l}
\bd{\F}_{i+1/2}^n = \dfrac{1}{2} \left(\bd{F_c} (\bd{W}_i^n)+\bd{F_c}(\bd{W}_{i+1}^n)\right) \,
- \,\dfrac{1}{2}\bd{\mathcal{D}}_{i+1/2}^n, \\
\end{array}
\end{equation*}
where $\bd{\mathcal{D}}_{i+1/2}^n$ is the numerical diffusion of the scheme.

Here we use here the framework of Polynomial Viscosity Methods (PVM) introduced in \cite{castro:2012} in order to define the numerical diffusion term. In particular, we use a generalization of the HLL scheme for non-conservative hyperbolic systems where
\begin{equation}\label{eq:difusion}
\bd{\mathcal{D}}_{i+1/2} = \a_0\,\left(\bd{\widehat{W}}_{i+1}-\bd{\widehat{W}}_i \right) \,+\, \a_1\,
\left(\bd{F_c} (\bd{W}_{i+1})-\bd{F_c}(\bd{W}_{i}) + \bd{\mathcal{S}}_{i+1/2}\right),
\end{equation}
with
\[\a_0 = \dfrac{S_R|S_L|-S_L|S_R|}{S_R-S_L},\hspace{0.2cm}\a_1 = \dfrac{|S_R|-|S_L|}{S_R-S_L},\]
being $S_L$ and $S_R$ approximations of the minimum and maximum wave speed. In practice,
$$
S_L= \min \left(u_{i}- \sqrt{g \cos \theta h_i}, \,  u_{i+1/2} -\sqrt{g \cos \theta h_{i+1/2}} \right),
$$
$$
S_R= \max \left(u_{i+1}+ \sqrt{g \cos \theta h_{i+1}}, \, u_{i+1/2} +\sqrt{g \cos \theta h_{i+1/2}} \right),
$$
where $h_{i+1/2},u_{i+1/2}$ are the Roe's averaged states.\\

One of the difficulties of this method is to ensure the well-balance property for \eqref{eq:steadysolution}. These steady states correspond to granular flows at rest for which the friction force is greater than pressure forces. To this aim, the numerical diffusion associated to the approximated Riemann solver must be zero in that case, in order to ensure $\p_t h = 0$ when the granular flow is at rest ($u=0$). Thus, we consider the reconstructed states

$$
\bd{\widehat{W}}_i = \left(\widehat{h}_{i+1/2}^-,h_{i+1/2}^-(u)_{i},0\right),
\qquad
\bd{\widehat{W}}_{i+1} = \left(\widehat{h}_{i+1/2}^+,h_{i+1/2}^+(u)_{i+1},0\right),
$$
in \eqref{eq:difusion}, where $\widehat{h}_{i+1/2}^\pm$ are defined as in \eqref{eq:recfondo}, taking in this case
\begin{equation}\label{eq:recfriccion1}
(\Delta Z_{i+1/2})_+ = \mbox{max}(0,z_{b,i+1}-z_{b,i} + \Delta {\mathcal C}_{i+1/2}),
\end{equation}
with $\Delta {\mathcal C}_{i+1/2} = -f_{i+1/2}\Delta x_{i+1/2}$ defined in terms of the Coulomb friction (see \cite{bouchut:2004}). We set
\begin{equation}\label{eq:recfriccion2}
f_{i+1/2} = -\mathop{proj}\limits_{g\cos\theta\mu(I_{|_b})}\left(\dfrac{-g\cos\theta(h_{i+1}  + z_{b,i+1} -h_i-z_{b,i})}{\Delta x} - \dfrac{u_{i+1/2}}{\Delta t}\right),
\end{equation}
where
\begin{equation}\label{eq:recfriccion3}
\mathop{proj}\limits_{g\cos\theta\mu(I_{|_b})}(X) = \left\{\begin{array}{lll}
X & \mbox{if} & |X| \leq g\cos\theta\mu(I_{|_b});\\
g\cos\theta\mu(I_{|_b}) \dfrac{X}{|X|} & \mbox{if} &  |X| > g\cos\theta\mu(I_{|_b}),
\end{array}\right.
\end{equation}
although other definitions of $f_{i+1/2}$ can be used (\cite{bouchut:2004}).

Once the numerical flux for the two first component is computed, the flux for the third one is
$$
\left[\bd{\F}_{i+1/2}^n\right]_{hw} = \left[\bd{\F}_{i+1/2}^n\right]_{h}\,w_{i+1/2}^{up,n},\quad\mbox{with}\quad w^{up}_{i+1/2} = \left\{
\begin{array}{lll}
w_i & \mbox{if} & \left[\bd{\F}_{i+1/2}\right]_{h}>0\\
w_{i+1} & \mbox{if} & \left[\bd{\F}_{i+1/2}\right]_{h}<0,
\end{array}
\right.
$$
where $\left[\bd{\F}_{i+1/2}^n\right]_{h}$, that approximates $(hu)_{i+1/2}$, denotes the first component of the numerical flux $\bd{\F}_{i+1/2}^n$.

Remark that following \cite{audusse:2004}, the scheme is positive preserving for the height thanks to the combination of the HLL scheme and the hydrostatic reconstruction. Moreover, the height will not be modified in the following two steps. Therefore, the complete scheme is positive preserving.

\subsection*{Step 2: Coulomb friction term}
In order to introduce the Coulomb friction term, we consider a semi-implicit scheme with an appropriate stopping criteria. From the physical point of view, the friction is a force that opposes the movement of the granular mass. When this friction is greater than the rest of the forces, then the flow must stop. The numerical treatment is based on this idea, which will be summarized in what follows. We refer the reader to  \cite{mangeney:2003,fernandezNieto:2008} for further details.

\medskip

We set  $h^{n+2/3}_i = h^{n+1/3}_i$, and define
$$
\begin{array}{l}
\displaystyle \sigma_{c,i}^n = \mu(I^n_{|_b})\left(g\cos\theta h_i^n + 2 q_i^n\right),\quad\mbox{with}\quad q_i^n = \left(q^n_{i-1/2} + q^n_{i+1/2}\right)/2, \\ \\
\left(hu\right)_i^{\star,n+1/3} = \left(hu\right)_i^{n+1/3} - \,\Delta t \left(h^{n}_i\left(\p_X q^{n}\right)_i + q_i^{n}\p_X\left(2b+h^{n}\right)_i \right)
\end{array}
$$
with 
$$
\left(\p_X q\right)_i = \dfrac{q_{i+1} - q_i}{\Delta x}, \quad
\left(\p_X\left(2b+h \right)\right)_i = \dfrac{\left(2b+h\right)_{i+1}-\left(2b+h\right)_{i-1}}{2\Delta x}.
$$
Then, the new values at this second step for the horizontal and vertical discharges are
$$
\left\{\begin{array}{llc}
\left(hu\right)^{n+2/3}_i = \left(hu\right)_i^{\star,n+1/3} - \Delta t\,\sigma_{c,i}^n \,SGN \left(\left(hu\right)_i^{\star,n+1/3}\right),   \\ & & \mbox{if}\quad \Delta t\,\sigma_{c,i}^n < \abs{\left(hu\right)_i^{\star,n+1/3}};\\[3mm]
\left(hw\right)_i^{n+2/3} = \left(hw\right)_i^{n+1/3}+ 2  \Delta t q_{i}^n,
\end{array}\right.
$$
otherwise
$$
\left(hu\right)_i^{n+2/3} =0, \quad \mbox{and} \quad \left(hw\right)_i^{n+2/3} = 0,
$$
where $SGN$ is the sign function.

\subsection*{Step 3: Non-hydrostatic pressure correction}
In the last step the non-hydrostatic effects are added using the momentum equations \eqref{eq:finalmodel_b},\eqref{eq:finalmodel_c} together with the incompressibility condition \eqref{eq:finalmodel_d}.

Taking into account system \eqref{eq:finalmodel}, we set $h^{n+1} = h^{n+2/3}=h^{n+1/3}$ and define
$$
\tilde{q}=q^{n+1}-q^n.
$$
Then we consider a projection method and we get

\begin{equation}
\label{eq:discrete_momentum_u}
\left(hu\right)^{n+1} = \left(hu\right)^{n+2/3} - \Delta t\,\Big( \p_X\left(h^{n+1} \tilde{q}\right) + 2 \tilde{q} \p_X b \Big),
\end{equation}
\begin{equation}
\label{eq:discrete_momentum_w}
\left(hw\right)^{n+1} = \left(hw\right)^{n+2/3} + 2\,\Delta t\,\tilde{q}.
\end{equation}
and the depth-averaged incompressibility equation
\begin{equation}
\label{eq:discrete_incomp}
\left(hw\right)^{n+1} = \left(hu\right)^{n+1}\p_X b - \dfrac{\left(h^2\right)^{n+1}}{2}\p_X u^{n+1}.
\end{equation}
The following elliptic problem is deduced for $\tilde{p}$,
\begin{eqnarray}
\label{eq:eliptic_pressure}
\left(h^2\right)^{n+1}\p_{XX} \tilde{q} &+& h^{n+1}\p_X h^{n+1}\p_X \tilde{q} + \Big(  h^{n+1}\p_{XX}\left(2b+h^{n+1}\right) - \left(\p_{X}\left(2b+h^{n+1}\right)\right)^2 - 4\Big) \tilde{q}  \nonumber\\[3mm]
&=& \dfrac{1}{\Delta t}\Big( 2\left(hw\right)^{n+1} - \left(hu\right)^{n+1}\p_{X}\left(2b+h^{n+1}\right) + h^{n+1}\p_{X}\left(hu\right)^{n+1}  \Big).
\end{eqnarray}
Finally, this equation is discretized in space at the interfaces $x_{i+1/2}$. Let us recall that the variables $(h)$, $(hu)$, and $(hw)$ are computed as averages in the control volumes, while $(q)$ is computed as point values at the interfaces. Therefore, we set
$$
h_{i+1/2} = \dfrac{h_i+h_{i+1}}{2},\qquad (hu)_{i+1/2} = \dfrac{(hu)_i+(hu)_{i+1}}{2},\qquad (hw)_{i+1/2} = \dfrac{(hw)_i+(hw)_{i+1}}{2},
$$
and we approximate of the derivative of the non-hydrostatic pressure deviation by
$$
\left(\p_{XX} \tilde{q}\right)_{i+1/2} = \dfrac{\tilde{q}_{i+3/2} - 2 \tilde{q}_{i+1/2} +\tilde{q}_{i-1/2}}{\Delta x^2} \qquad
\left(\p_X \tilde{q} \right)_{i+1/2} = \dfrac{\tilde{q}_{i+1/2}-\tilde{q}_{i-1/2}}{2\Delta x}.
$$
Moreover, we set
$$
\left(\p_X h\right)_{i+1/2} = \dfrac{h_{i+1}-h_{i}}{\Delta x}, \qquad \left(\p_X b\right)_{i+1/2} = \dfrac{b_{i+1}-b_{i}}{\Delta x},
$$
and
$$
\left( \p_{XX} \left( 2b+h\right)\right)_{i+1/2} = minmod\left(\Delta_{2b+h}^+,\Delta_{2b+h}^c,\Delta_{2b+h}^-\right),
$$
where
$$\Delta_{2b+h}^+ = \dfrac{\p_{X} \left( 2b+h\right)_{i+3/2} - \p_{X} \left( 2b+h\right)_{i+1/2}}{\Delta x}, \quad
\Delta_{2b+h}^- = \dfrac{\p_{X} \left( 2b+h\right)_{i+1/2} - \p_{X} \left( 2b+h\right)_{i-1/2}}{\Delta x},$$
and
$\Delta_{2b+h}^c = \left(\Delta_{2b+h}^+ + \Delta_{2b+h}^-\right)/2$. \\\\

\noindent Then, a tridiagonal linear system is obtained for the unknown values $\{ \tilde{q}_{i+1/2} \}_i$.
Once this linear system is solved, the values of $\left\{ ( \, (hu)_i^{n+1}, (hw)_i^{n+1},q_{i+1/2}^{n+1} \, ) \right\}_{i\in\mathcal{I}}$ are updated using \eqref{eq:discrete_momentum_u} and \eqref{eq:discrete_momentum_w}, leading to
$$
\begin{array}{l}
\displaystyle \left(hu\right)_i^{n+1} = \left(hu\right)_i^{n+2/3} \,-\,\Delta t \left(h^{n+1}_i\left(\p_X \tilde{q} \right)_i + \tilde{q}_i\p_X\left(2b+h^{n+1}\right)_i \right),  \\ \\
\displaystyle \left(hw\right)_i^{n+1} = \left(hw\right)_i^{n+2/3} \,+\,2\,\Delta t \tilde{q}_i,  \\ \\
q_{i+1/2}^{n+1}=\tilde{q}_{i+1/2}+q_{i+1/2}^n,
\end{array}
$$
with $$\tilde{q}_i = \dfrac{\tilde{q}_{i-1/2}+\tilde{q}_{i+1/2}}{2},\qquad \left(\p_X \tilde{q}\right)_i = \dfrac{\tilde{q}_{i+1} - \tilde{q}_i}{\Delta x}, \quad
\left(\p_X\left(2b+h\right)\right)_i = \dfrac{\left(2b+h\right)_{i+1}-\left(2b+h\right)_{i-1}}{2\Delta x}.
$$

\section{Numerical test}\label{se:tests}
In this section, we present some numerical tests in order to validate the non-hydrostatic model and the numerical approach introduced in this paper. Comparisons with a hydrostatic version of the proposed model (see Remark \ref{re:remark_modelo_hidrostatico}) will be shown.

First, we study the influence of the choice of the coordinate system (local or Cartesian) when using the hydrostatic and the non-hydrostatic model. In a second series of tests, we compare with experimental data of granular collapse over inclined planes described in \cite{mangeney:2010}. Comparisons are carried out using both the hydrostatic and non-hydrostatic models in local coordinates.

Notice that, in this section, whenever we speak about hydrostatic/non-hydrostatic model in local or Cartesian coordinates, we refer to the direction along which the shallowness approximation is applied and the depth-average procedure is performed starting from the $2$D Navier-Stokes system. This direction is normal to the reference plane $\wtd{b}$ for local coordinates and the vertical $z$-direction for the Cartesian coordinates.

All the simulations are carried out with a constant mesh size, $\Delta x = 0.5$ cm, and an adaptive time step, $\Delta t$, computed with $CFL = 0.5$. The rheological parameters of the granular material are given in Table \ref{tab:tabla_datos}. 

\begin{table}[!ht]
	\begin{center}
		\begin{tabular}{c|c|c|c|c}
			$d_s$ (mm)  & $\mu_s$ & $\mu_2$ & $I_0$ & $\varphi_s$\\
			\hline			
			$0.7$ & $\tan(25.5^\circ)$ & $\tan(36^\circ)$ & $0.279$ & $0.62$\\
		\end{tabular}
		\caption{\it{Rheological parameters considered in all the numerical tests.}}
		\label{tab:tabla_datos}
	\end{center}
\end{table}

\subsection{Influence of the coordinate system}
In this test, we first propose to analyze how much the use of local coordinates is important. To do so, let us compare the results obtained when one uses local
 or Cartesian coordinates. The simulations will be performed using the non-hydrostatic model presented here as well as its hydrostatic counterpart. For the sake of simplicity, we consider that the bottom is defined by the reference slope plane. In order to compute the simulation corresponding to system \eqref{eq:finalmodel} in local coordinates we must set $\wtd{b}_{loc}(x) = -\tan\theta\left(x-x_{end}\right)$ and $b_{loc} = 0$.  Conversely,  in Cartesian coordinates we have to define $b_{Cart}(x) = -\tan\theta\left(x-x_{end}\right)$, $\wtd{b}_{Cart} = 0$, and write $g$ instead of $g\cos\theta$ everywhere in system \eqref{eq:finalmodel}.

Let us remark that in this case the term $2\wb{q}\p_Xb$ in the non-hydrostatic model \eqref{eq:finalmodel_b} vanishes when the model is written in local coordinates, whereas it is equal to $-2\wb{q}\tan\theta $ in the Cartesian version of the model. This behavior is very different if we compare with the hydrostatic model, where this term is always zero.

We shall analyze the influence of the choice or coordinates by considering a numerical test where the initial condition is well defined in both coordinate systems.
\begin{figure}[!b]
	\begin{center}
		\includegraphics[width=0.85\textwidth]{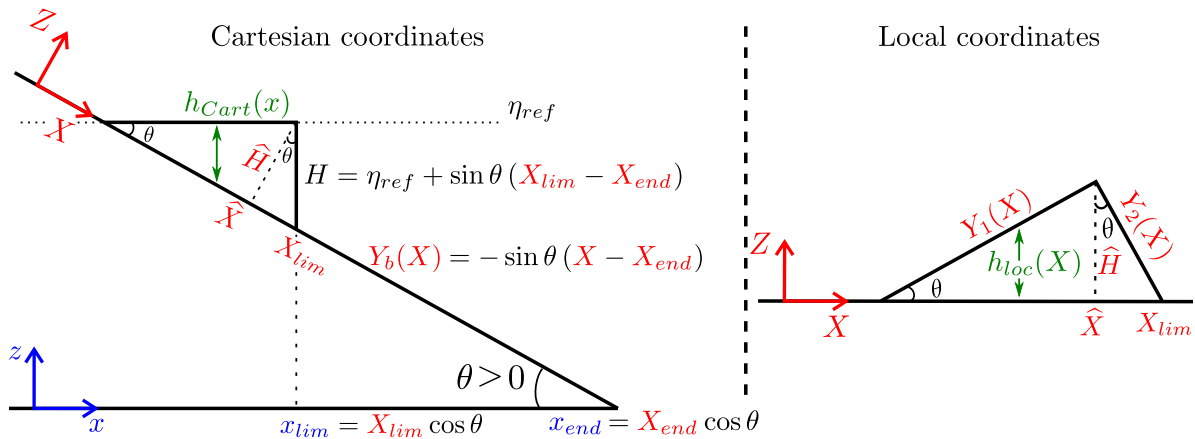}
	\end{center}
	\caption{\label{fig:local_cart_esq} \it{Sketch of the initial condition in Cartesian (blue) and local (red) coordinates.}}
\end{figure}
We consider a computational domain given by the interval $[x_0,x_{end}]$. The initial condition is shown in Figure \ref{fig:local_cart_esq}. We shall denote by $h_{loc}(X)$ and $h_{Cart}(x)$ the initial height function in local and Cartesian coordinates respectively. The initial height in Cartesian coordinates is given by
$$h_{Cart}(x) = \max\left(\eta_{ref} + \tan\theta \left(x-x_{end}\right),0\right), \quad\mbox{for  } x<x_{lim},$$ where $\eta_{ref}$ is a reference level, $x_{end}$ is the right boundary of the computational domain and $x_{lim}$ is the initial front position. \\

In order to define the initial condition using local coordinates we use that $x = X\cos\theta$. Then, $X_{lim}=x_{lim}/\cos \theta$ is the initial position of the front position in local coordinates. We also use that the distance from 
$Y_b(X_{lim})$ to the reference level (vertically measured) is $$H = \eta_{ref} + \sin\theta\left(X_{lim}-X_{end}\right),$$ which is the maximum height of the flow when considering Cartesian coordinates. Then, the maximum height of the initial condition defined in local coordinates is $\wht{H} = H\cos\theta$, located in $\wht{X} = X_{lim} - \wht{H}\tan\theta$ (see figure \ref{fig:local_cart_esq}b). Using this notation we can define the initial condition in local coordinates as follows:
$$h_{loc}(X) = \max(h_1(X),0), \text{  with }
h_1(X) = \left\{\begin{array}{lll}
y_1(X) = \max\left(\wht{H} + \tan\theta\left(X-\wht{X}\right),0\right), & \mbox{if}&X\leq\wht{X},\\
y_2(X) = \max\left(\dfrac{1}{\tan\theta}\left(X_{lim}-X\right),0\right), & \mbox{if}&X>\wht{X}.
\end{array}\right.
$$

\begin{figure}[!b]
	\begin{center}
		\includegraphics[width=0.49\textwidth]{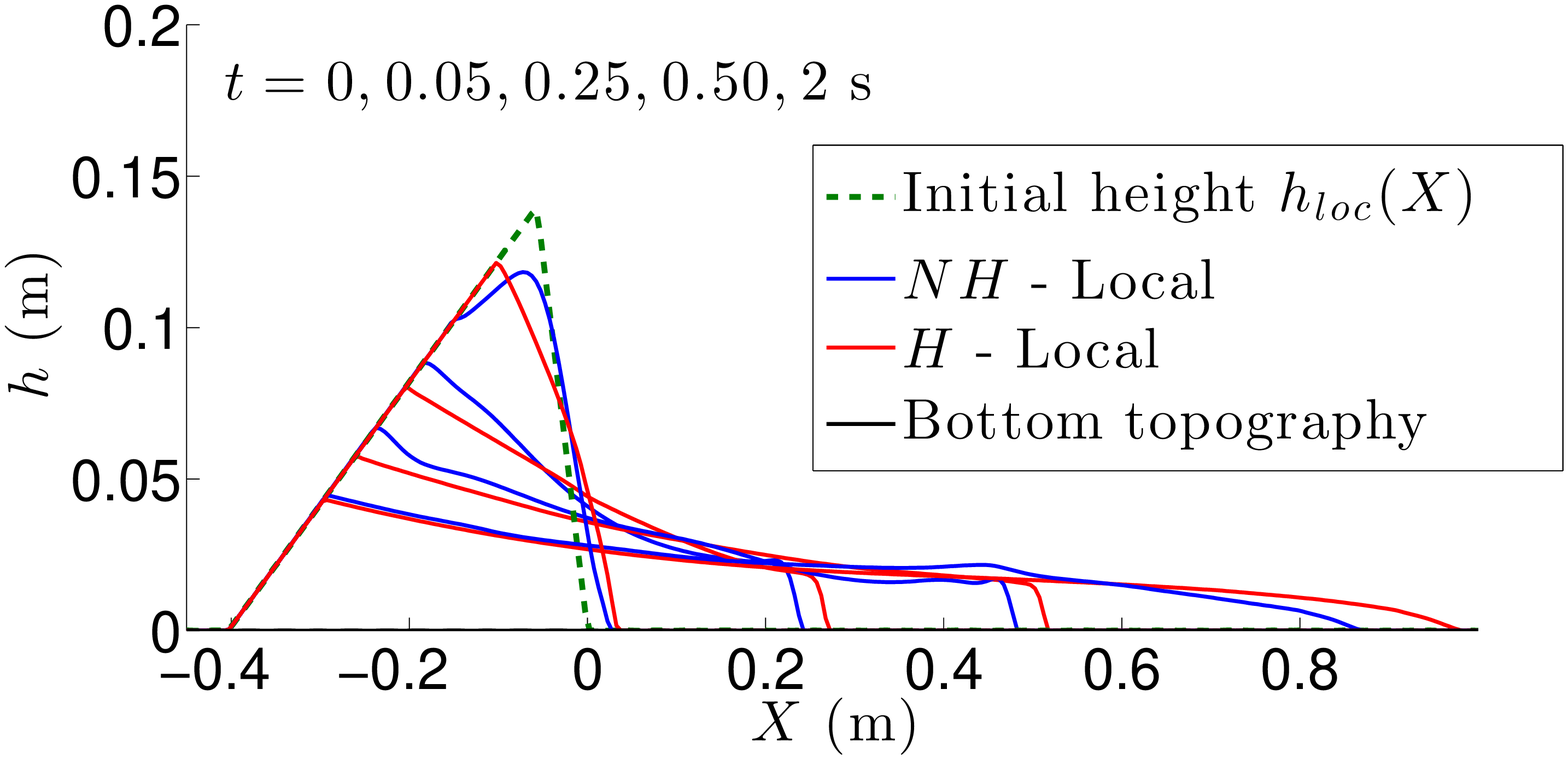}
	\end{center}
	\caption{\label{fig:local_cart_local} \it{Deposits at different times of the hydrostatic (red lines) and the non-hydrostatic (blue lines) models computed in local coordinates. Dashed green line corresponds to the initial height.}}
\end{figure}

In practice, we set the slope $\theta = 22^{\circ}$ and the computational domain (in local coordinates) by $X\in [-0.5,2.7]$. At initial time, the considered granular mass is at rest, the position of the front is assumed to be at $X=0$, i.e. $X_{lim} = 0$, and the maximum (local) height is assumed to be $\wht{H} = 0.14$ m. This is equivalent to consider a reference level $\eta_{ref} = 0.14/\cos\theta + x_{end}\tan\theta$ m.

\begin{figure}[!ht]
	\begin{center}
		\includegraphics[width=0.49\textwidth]{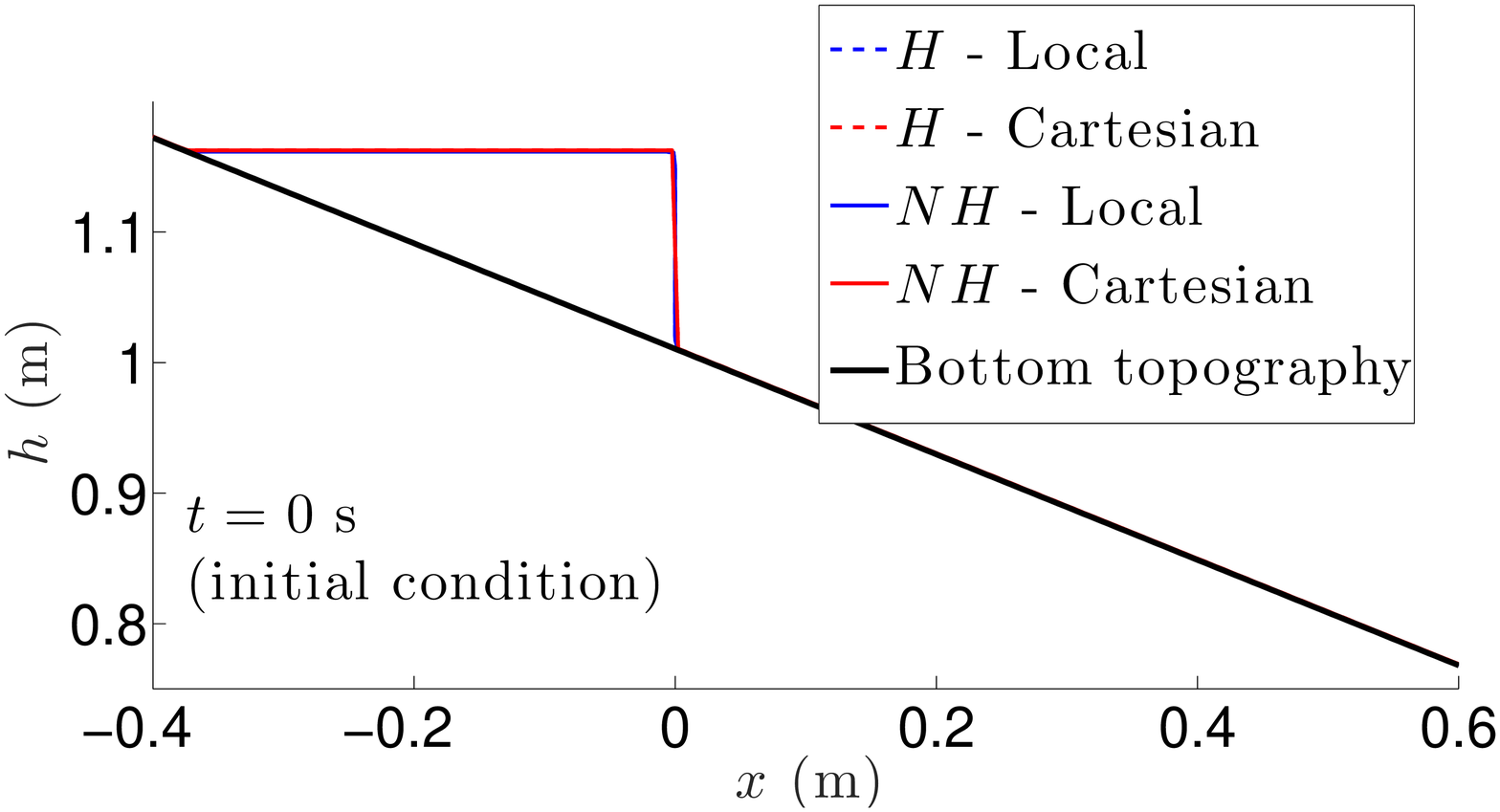}
		\includegraphics[width=0.49\textwidth]{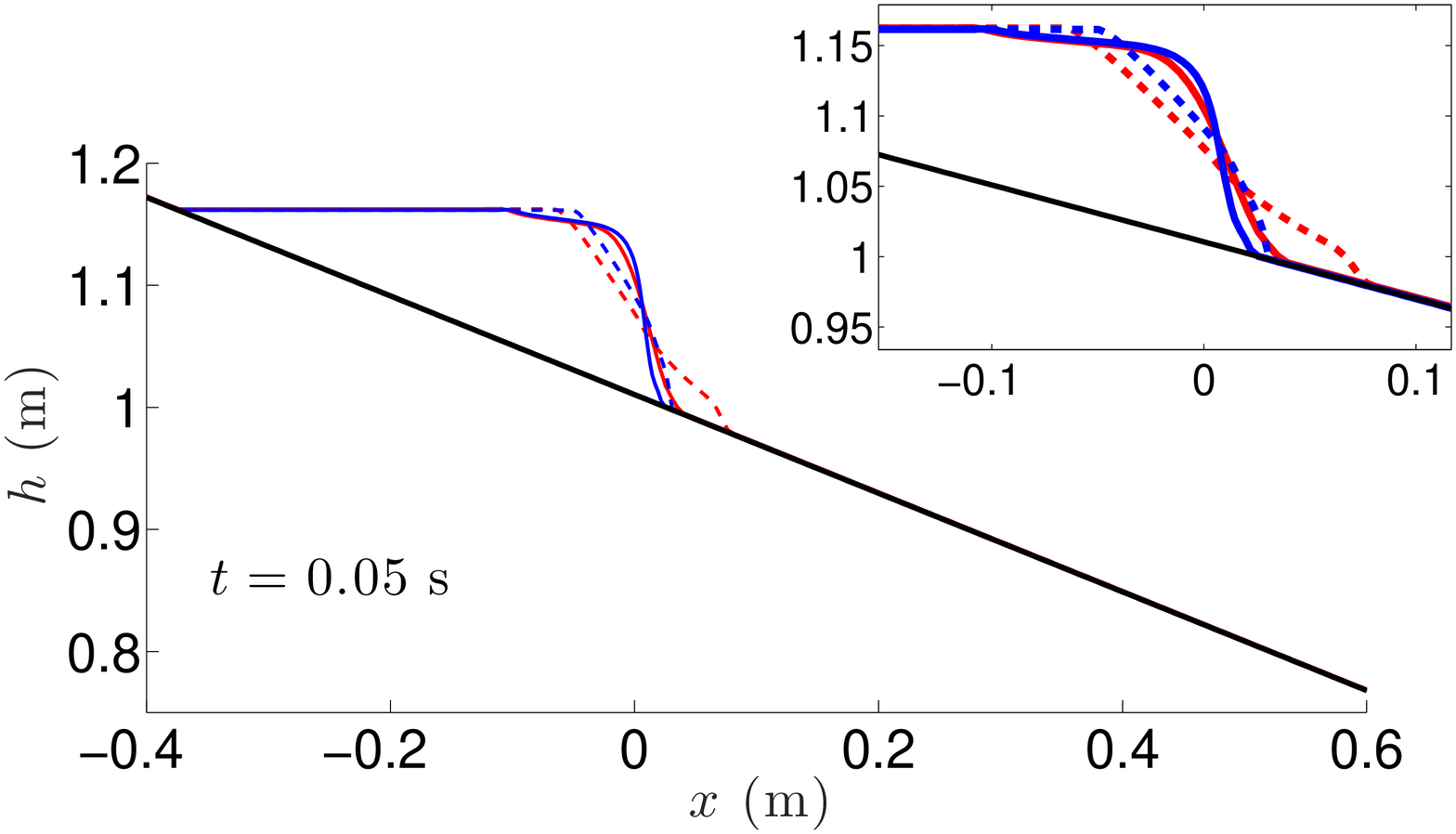}
		\includegraphics[width=0.49\textwidth]{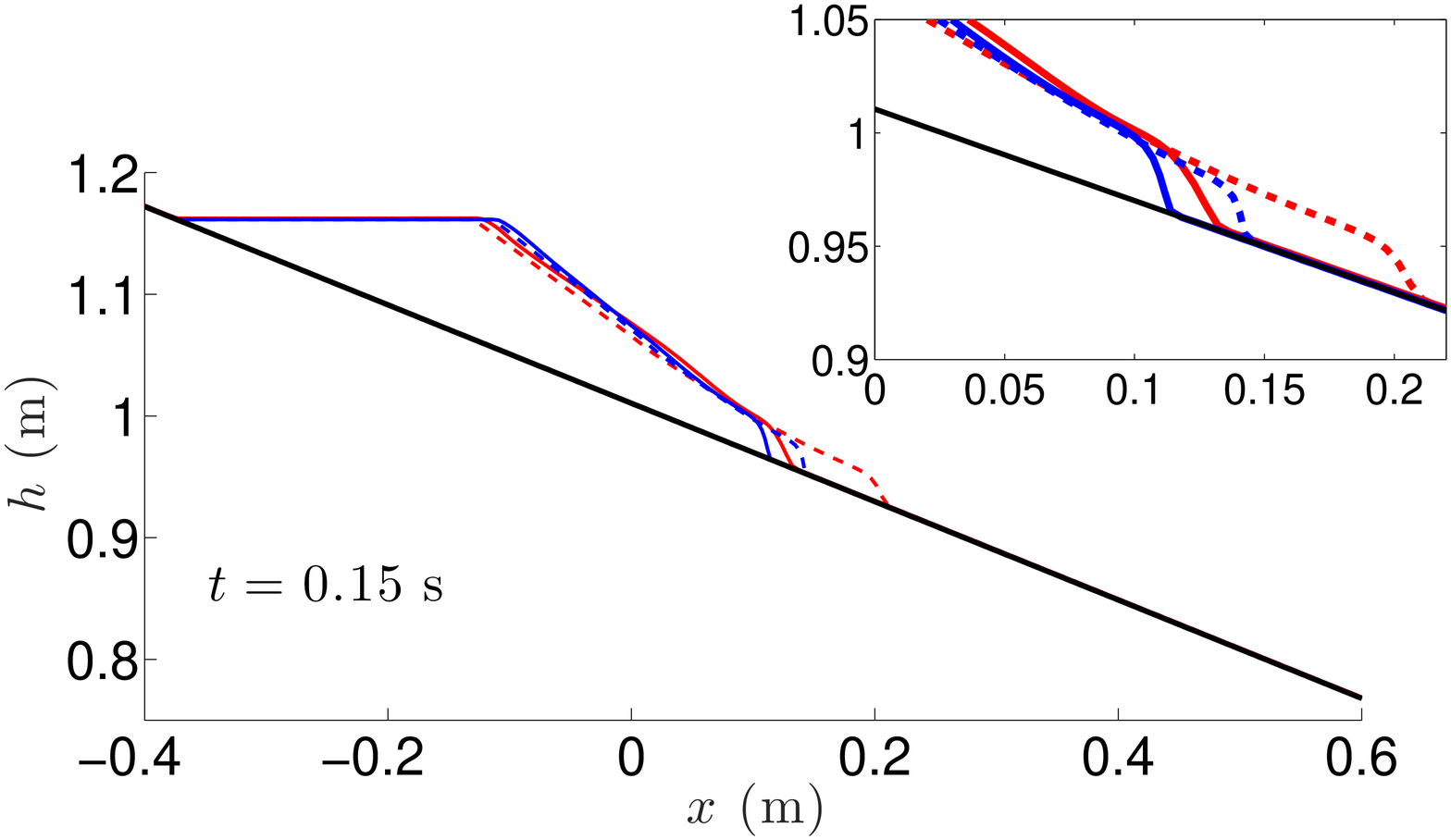}
		\includegraphics[width=0.49\textwidth]{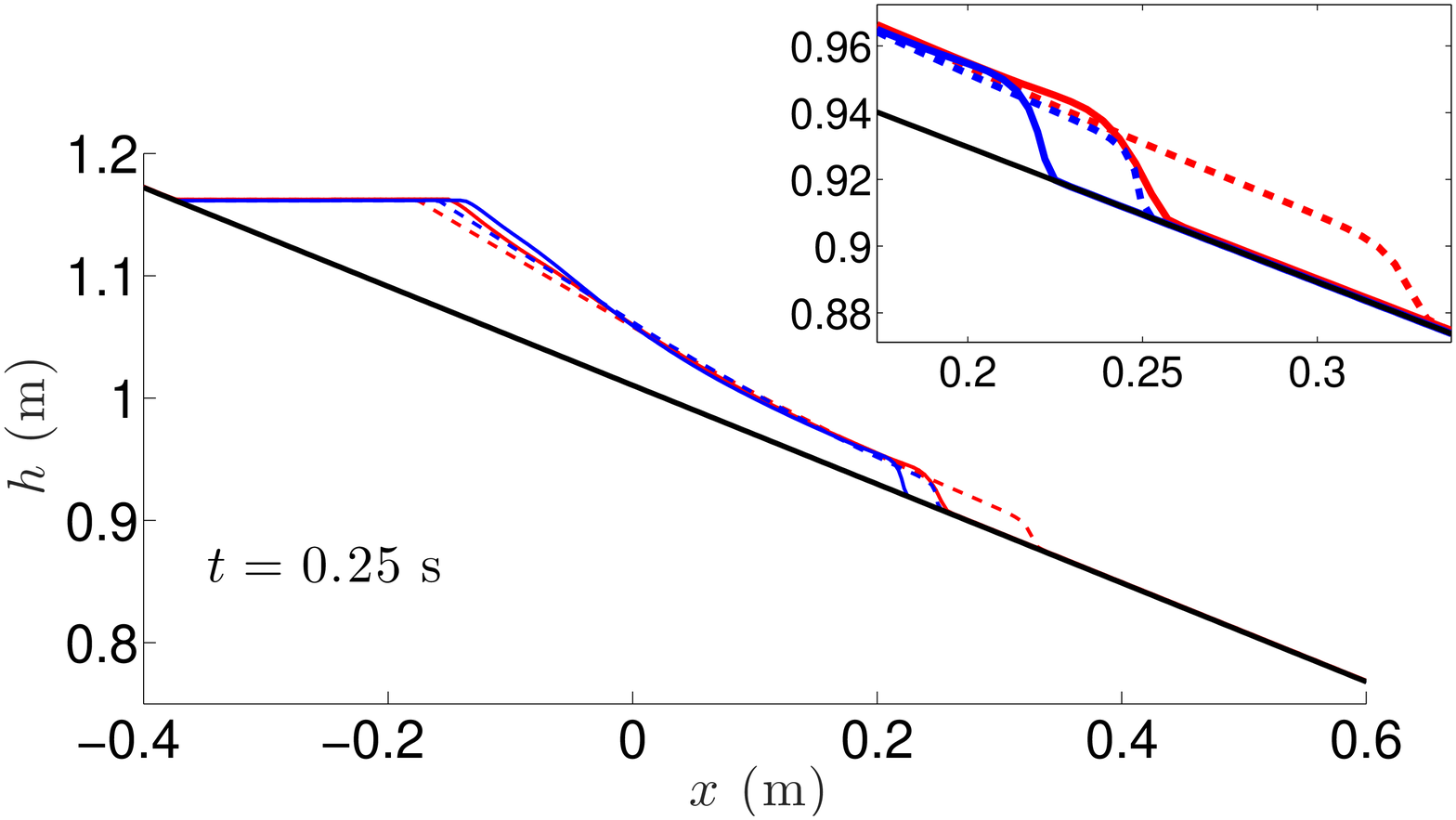}
		\includegraphics[width=0.49\textwidth]{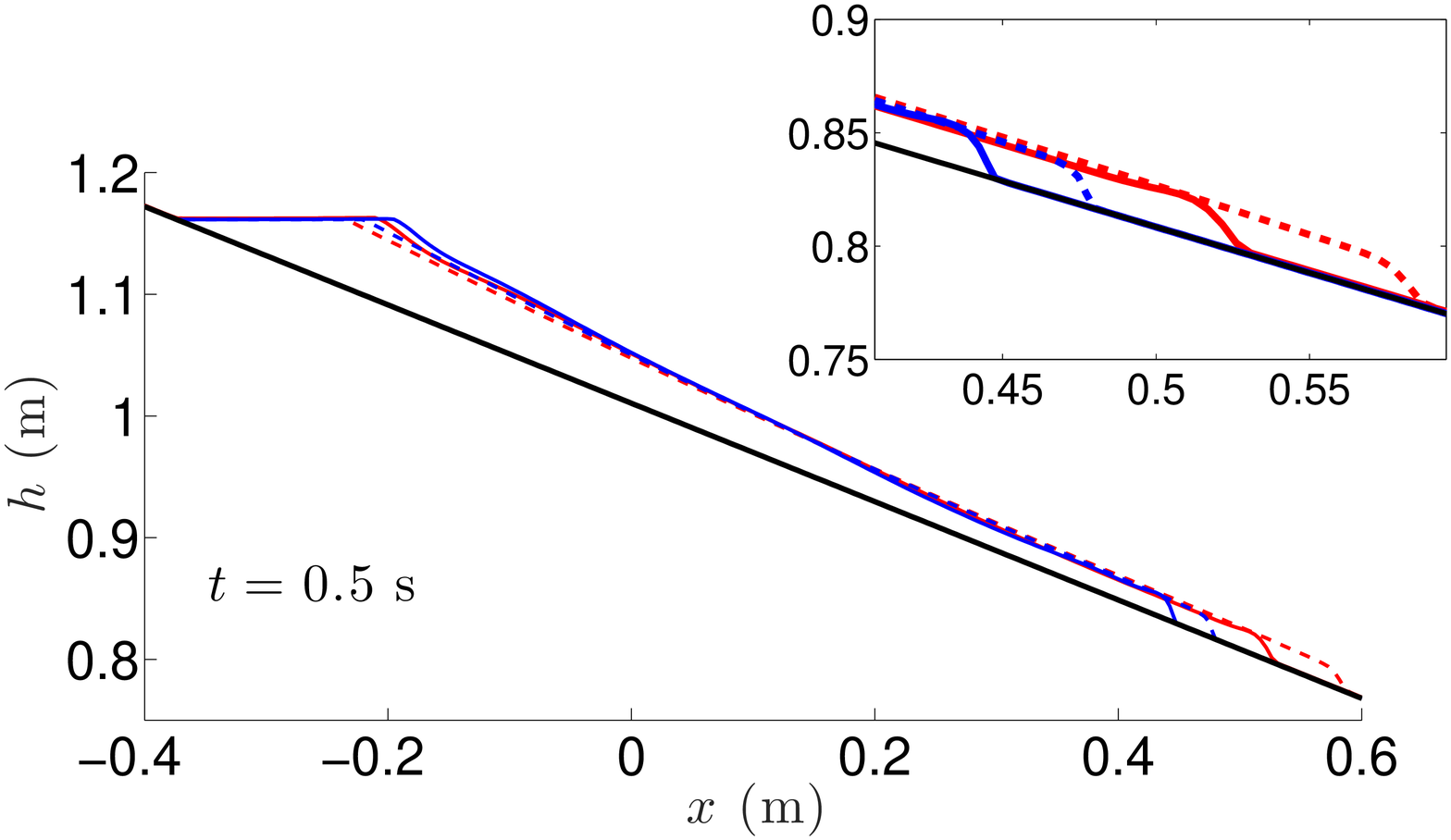}
		\includegraphics[width=0.49\textwidth]{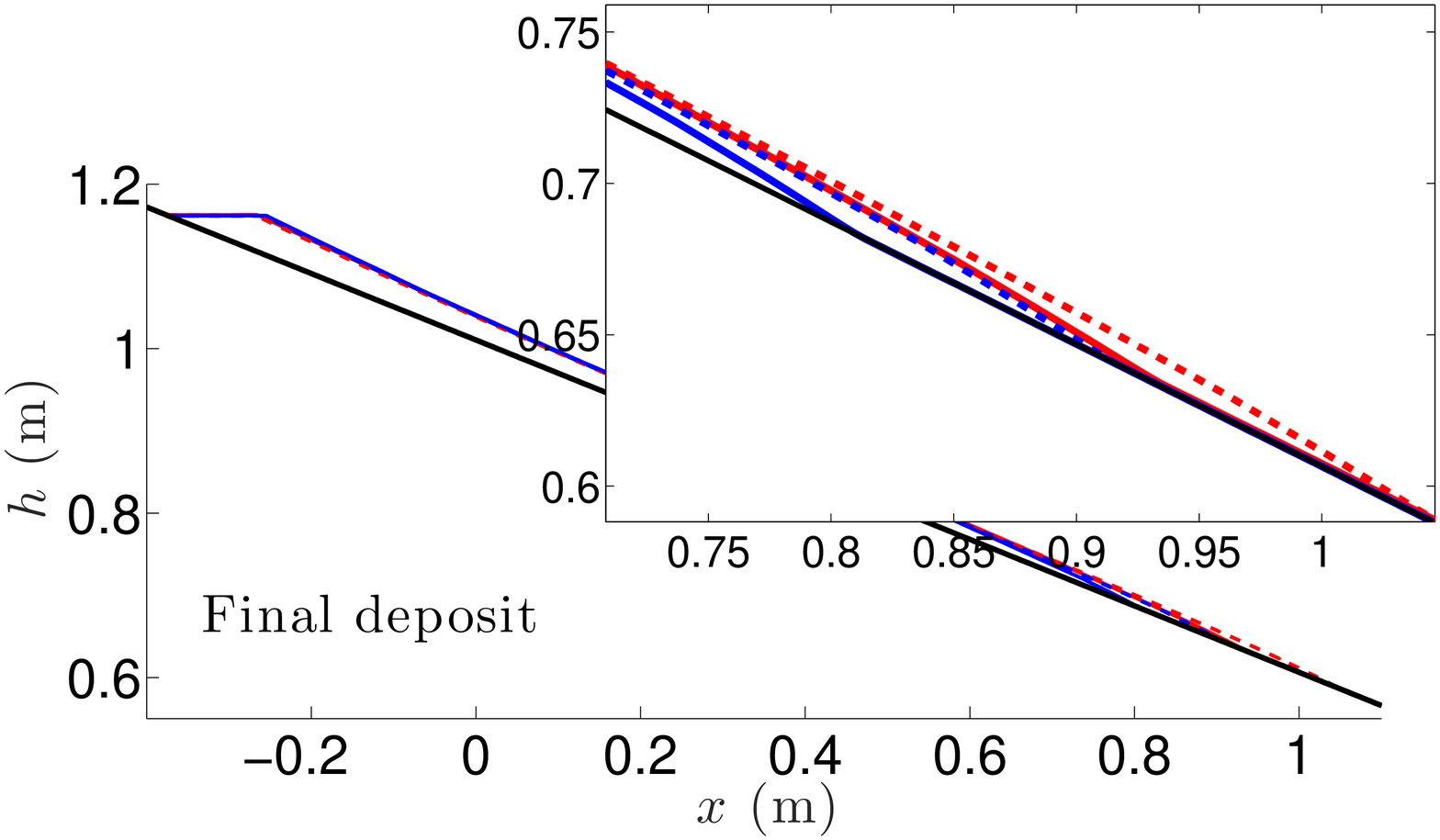}
	\end{center}
	\caption{\label{fig:local_cart} \it{Thickness profiles of the granular mass flowing on a plane of inclination $\theta=22^{o}$ at different times. Blue lines correspond to local models with hydrostatic (dashed) and non-hydrostatic (solid) pressure, while red lines are the solutions of Cartesian models with hydrostatic (dashed) and non-hydrostatic (solid) pressures. Inset figures show zooms of the front position.} }
\end{figure}

\begin{figure}[!ht]
	\begin{center}
		\includegraphics[width=0.49\textwidth]{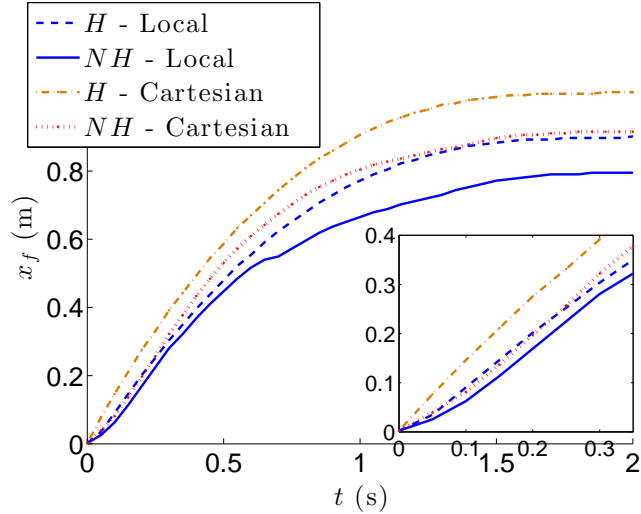}
	\end{center}
	\caption{\label{fig:local_cart_runout} \it{Time evolution  of the granular front position $x_f$ computed with the hydrostatic and the non-hydrostatic models in local coordinates (dashed and solid blue line, respectively), and the hydrostatic and the non-hydrostatic models in Cartesian coordinates (dot-dashed brown and dotted red line, respectively). Inset figure shows a zoom at short times.}}
	\end{figure}

	\begin{figure}[!ht]
		\begin{center}
			\includegraphics[width=0.49\textwidth]{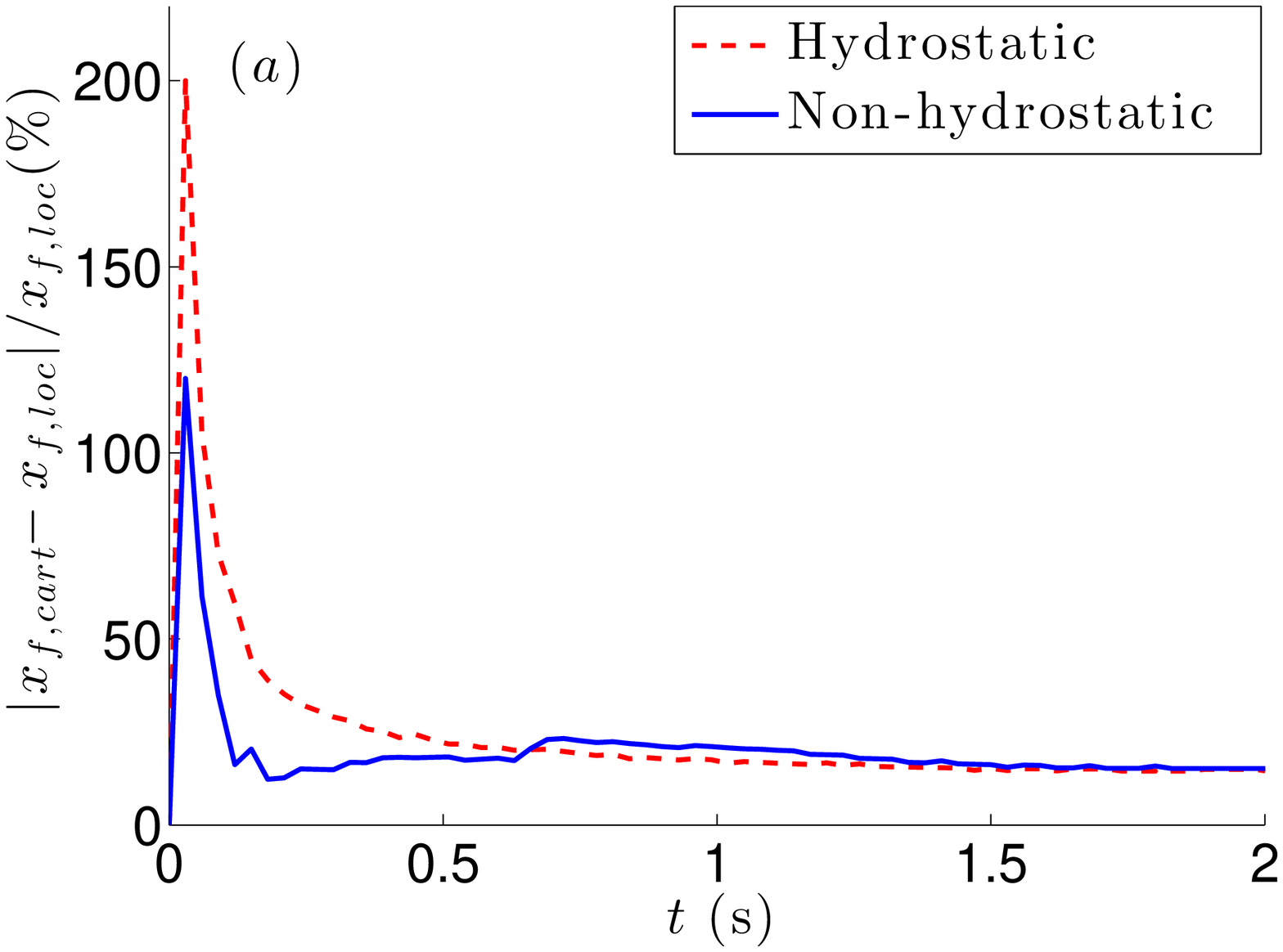}\\
			\includegraphics[width=0.49\textwidth]{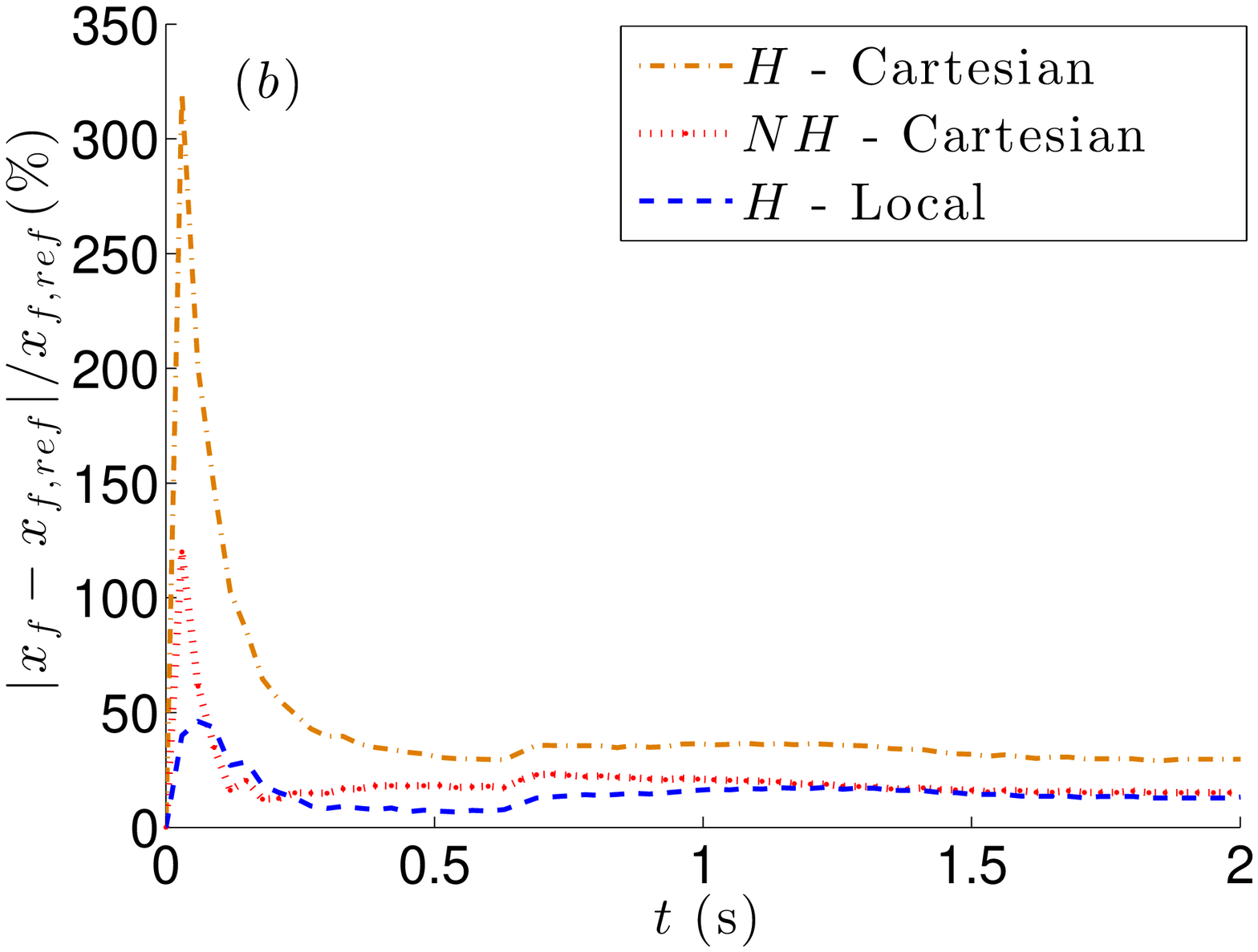}
			\includegraphics[width=0.49\textwidth]{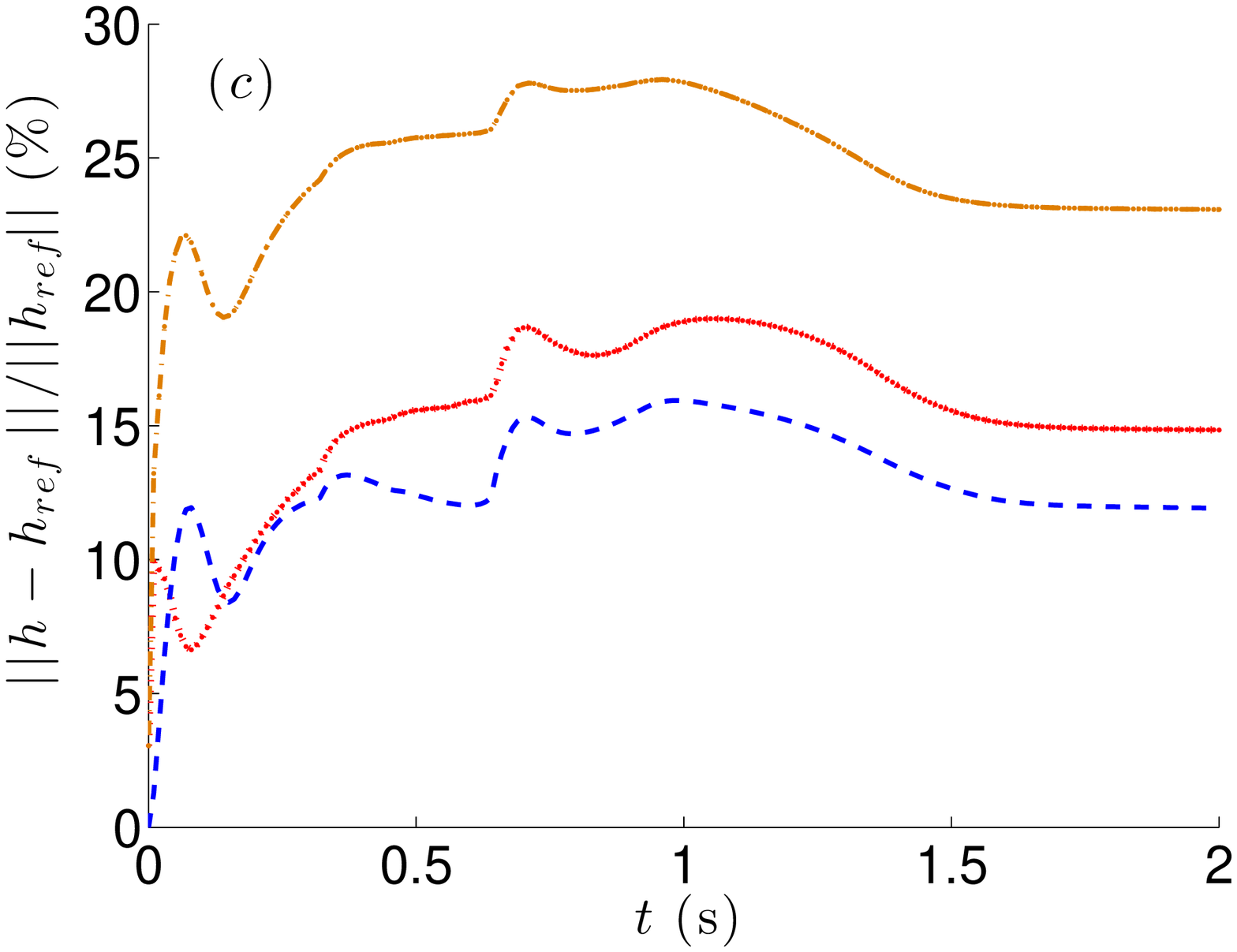}			
		\end{center}
		\caption{\label{fig:local_cart_error_runout} \it{Time evolution of the relative errors of (a) the front position $x_f$ computed with the local hydrostatic (H) model (dashed blue line) the Cartesian NH and the H models (dotted red line and dot-dashed brown line respectively) taking as reference the runout obtained with the NH model in local coordinates; (b) runout (c) granular mass thickness computed and with local and Cartesian coordinates for the non-hydrostatic models (blue line) and the hydrostatic models (dashed red line).}
		}
	\end{figure}

Figure \ref{fig:local_cart_local} shows the time evolution of hydrostatic (H) and non-hydrostatic (NH) local models in a local view, i.e. the flow thickness $h$ is represented in the direction normal to the slope. The spreading of the granular mass simulated with the non-hydrostatic model is slower than with the hydrostatic model. As a result the front position is always located further downslope with the hydrostatic model, leading to significantly longer runout distance. Concretely, it is $12.7\%$ bigger using the hydrostatic model. The maximum thickness of the flowing mass and of the deposit is also lower with the hydrostatic model, except at the very beginning of the flow. In Figure \ref{fig:local_cart} we show the evolution of the flowing mass obtained with the hydrostatic and the non-hydrostatic models, in both coordinate systems. The simulations obtained with the hydrostatic model, both in local and Cartesian coordinates, are faster during the first instants than non-hydrostatic models. The non-hydrostatic model in Cartesian coordinates generate faster flows than the hydrostatic model in local coordinates for $t> 0.22$ s, approximately, leading to larger travelling distances of the granular front and consequently larger runout distance. From that time, the two models in Cartesian coordinates go further than the models in local coordinates. One of the outcome of this comparison is that the non-hydrostatic Cartesian model does not give the same results as the local hydrostatic model, contrary to what was assumed in \cite{denlinger:2004}. This is also shown in Figure \ref{fig:local_cart_runout}, where we see the time evolution of the granular front position. In the inset figure we see that the front position simulated with the non-hydrostatic (NH) model in Cartesian coordinates is slightly smaller than the one computed with the hydrostatic (H) model in local coordinates at short times while it is higher later on, as commented before. The final runout distance using the NH Cartesian model and the H local model are however similar as also assumed in \cite{denlinger:2004}. Note also that the time change of the front position simulated with the NH local model exhibit a curvature change during the first instants as observed in laboratory experiments (Figure 9a of \cite{mangeney:2010}) while is is not the case with hydrostatic models.

It is well-known that models in local coordinates are more appropriate than model in Cartesian coordinates, since local models compute the velocity in the direction tangent to the topography, which is the relevant direction for these dense granular flows. As a result, we will calculate the error made when using Cartesian coordinates instead of local coordinates. In the same way, we will chose as a reference the NH local model for which the shallow approximation and depth-integration is performed in the good direction and that includes some non-hydrostatic contribution.

In Figure \ref{fig:local_cart_error_runout}a we show the relative error of the front position between the results obtained in Cartesian coordinates compared to the local coordinates for both the hydrostatic and non-hydrostatic models. We can observe that hydrostatic models are more dependent on the coordinate system. That is an expected behavior as fully $3$D non-hydrostatic results are independent of the choice of the coordinate system.
In Figure \ref{fig:local_cart_error_runout}b and \ref{fig:local_cart_error_runout}c we show relative errors on the front position and height along the domain computed with the different models compared to the reference solution obtained with the NH local model. As expected, the errors corresponding to the solution of the Cartesian hydrostatic model are the biggest one, reaching 320$\%$ during the first instants up to 30$\%$ on the runout distance. These errors are lower when looking at the height along the domain. This error is greater also for the H Cartesian model, being approximately 23$\%$ at final time, whereas it is 15$\%$ and 12$\%$, approximately, for the NH Cartesian and H local models, respectively. This behavior is also seen in Figure \ref{fig:local_cart_runout}.
We can conclude that hydrostatic models in Cartesian coordinates predict a much too long runout distance. Nevertheless, and interestingly, the Cartesian non-hydrostatic model and the local hydrostatic model give similar deposits even though the dynamics is different, as shown in figures \ref{fig:local_cart_error_runout} and \ref{fig:local_cart_runout}. This partly supports the assumption of \cite{denlinger:2004} but only for the deposit. Indeed, these authors proposed a hydrostatic model in Cartesian coordinates with a correction of the pressure accounting for an approximation of the vertical acceleration. They showed that their model produces similar results to the ones obtained with the hydrostatic local model for the analytical solution of a dam break problem (see their Figure 4b).

This last result motivates the next test, where these two models (H-local and NH-Cartesian) are compared in a more general case, with a more complex topography.

\subsection{Hydrostatic local model vs Non-hydrostatic Cartesian model}\label{se:tests_HvsNH}

The goal of this test is to show a qualitative comparison of the hydrostatic local and the non-hydrostatic Cartesian model for flows on a complex topography. We consider here a granular mass, with the same rheological properties (see Table \ref{tab:tabla_datos}) as in previous test, in the computational domain $[-3,3]$. In this case we take $300$ nodes for the horizontal discretization.
The topography, in Cartesian coordinates, is given by $\wtd{b}_{Cart} = 0$ and
\begin{equation}
\label{eq:batemetria_test2}
b_{Cart}(x) = 1-\tanh(x)  + 0.3e^{-10(x-1)^2}+ 0.5e^{-10(x-3)^2},
\end{equation}
and the initial height is 
$$h_{Cart}(x) = \max(\eta(x)-b_{Cart}(x),0), \quad\text{ with }  \quad
\eta(x) = \left\{\begin{array}{lll}
y_0-1+e^{-0.5(x-x_0)^8}, & \mbox{if}&x\leq 0,\\
0, & &\mbox{otherwise},
\end{array}\right.
$$
with $x_0 = -1.53$ m and $y_0 = 1.91$ m.
\begin{figure}[!ht]
	\begin{center}
		\includegraphics[width=0.69\textwidth]{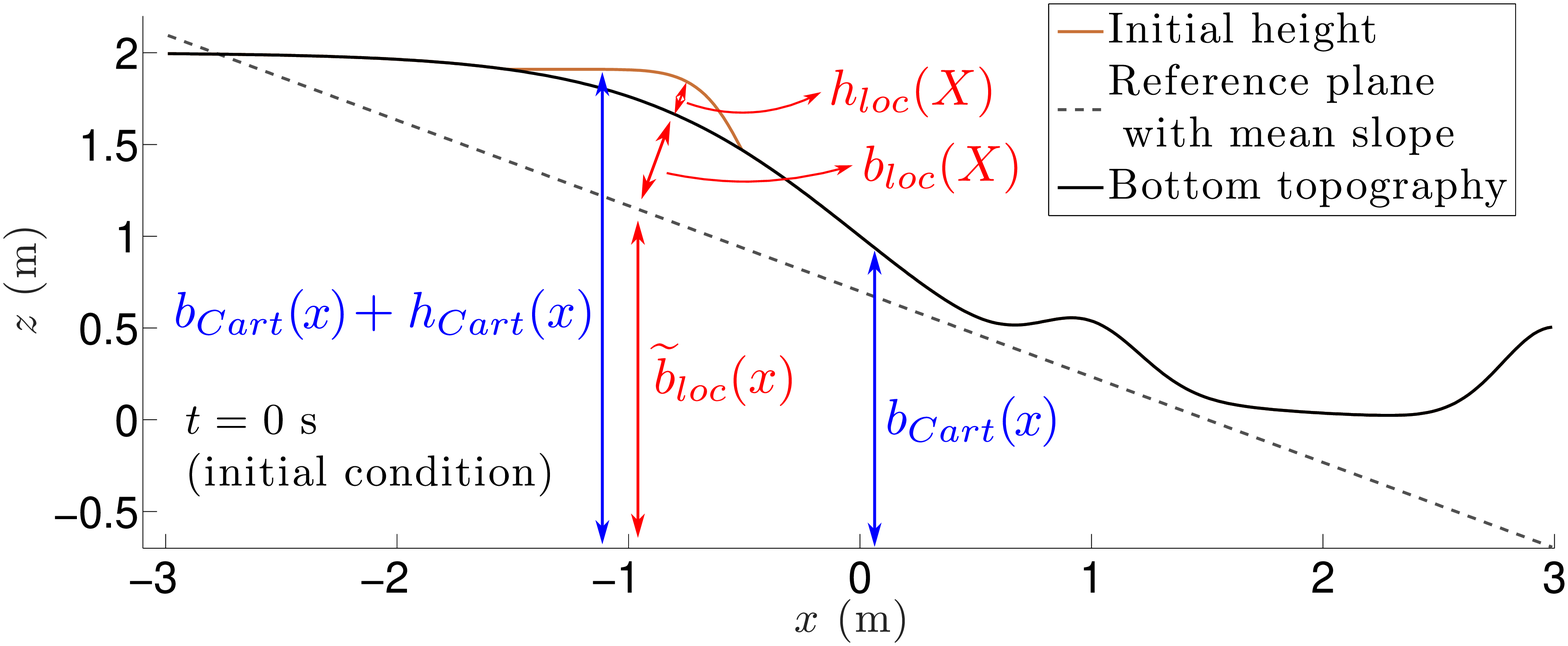}
	\end{center}
	\caption{\label{fig:test2_ini} \it{Initial height (solid red line), bottom topography (solid black line) and reference plane with the mean slope used for the local model (dashed gray line).}}
\end{figure}

Defining this initial configuration in a local coordinates is not a simple task. First, a reference plane $\wtd{b}_{loc}(x)$, whose slope is the mean slope of the topography, is defined. In our case, $\wtd{b}_{loc}(x)=0.7-\tan(25^\circ)x$. Then, the topography $b_{loc}(X)$ is defined as the distance from $\wtd{b}_{loc}(x)$ to $b_{loc}(X)$, measured in the normal direction to the reference plane $\wtd{b}_{loc}$ (see Figure \ref{fig:test2_ini}). Analogously, the height $h_{loc}(X)$ is the distance from $b_{loc}(X)$ to $h_{loc}(X)$. The granular mass is supposed to be initially at rest. After some time the grains stop, leading to three separate regions of material at rest.

\begin{figure}[!ht]
	\begin{center}
		
		\includegraphics[width=0.49\textwidth]{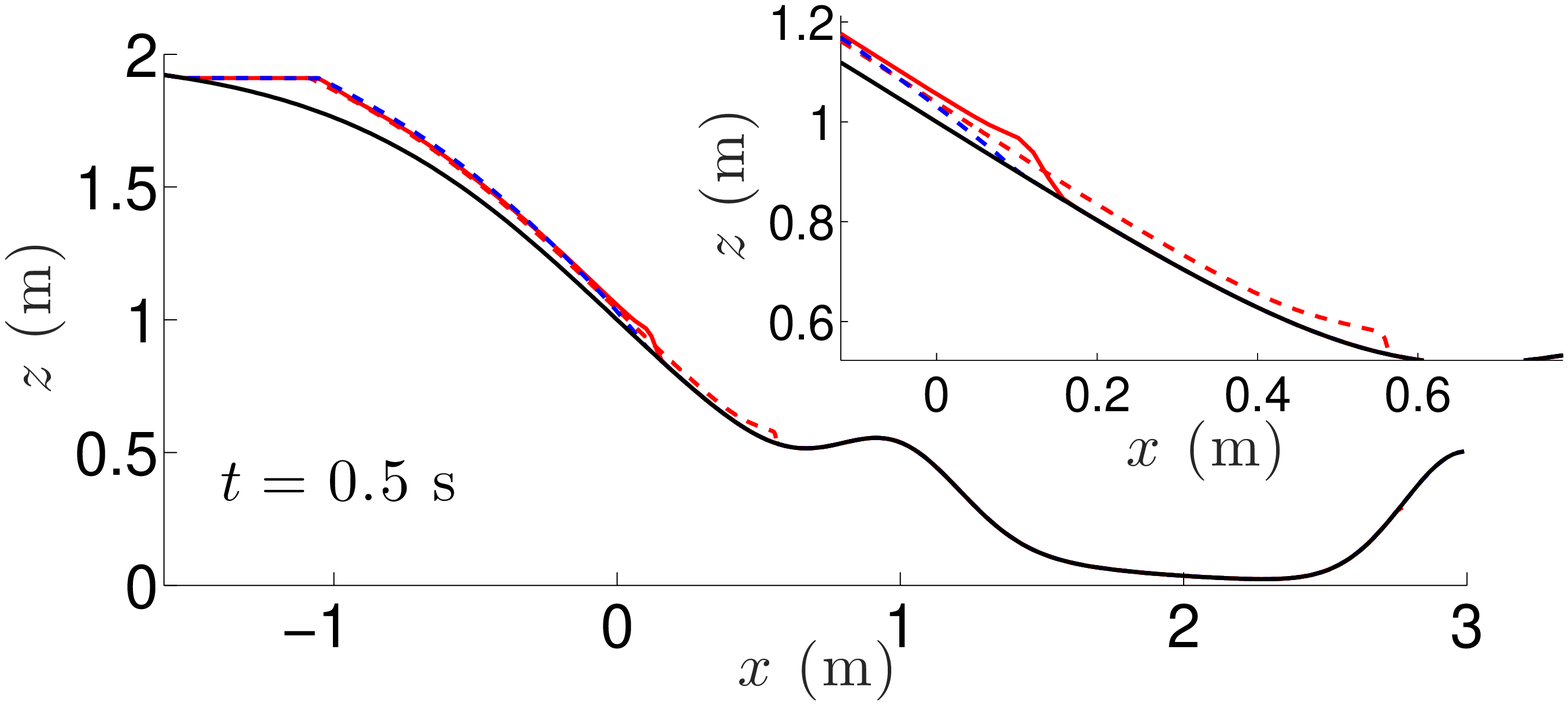}
		\includegraphics[width=0.49\textwidth]{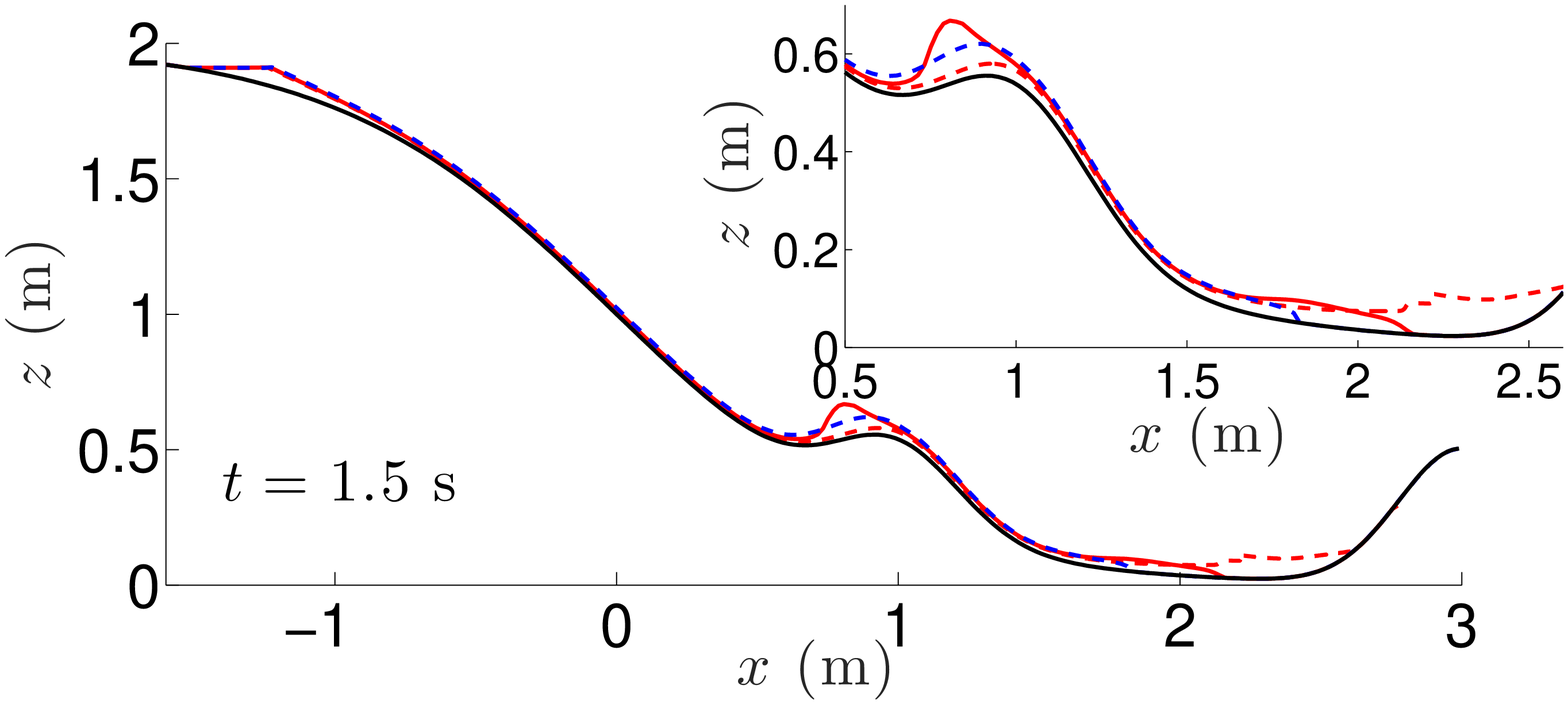}
		\includegraphics[width=0.49\textwidth]{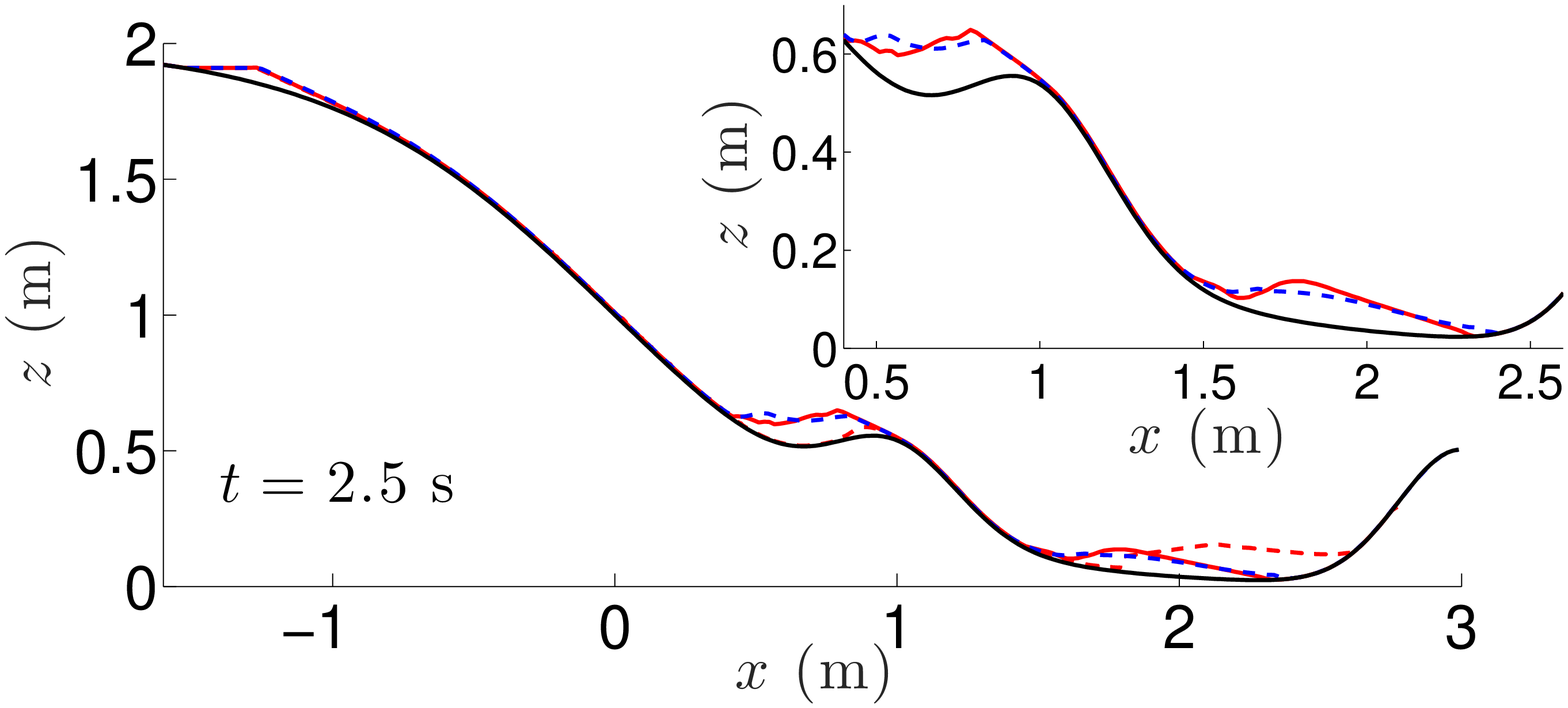}
		\includegraphics[width=0.79\textwidth]{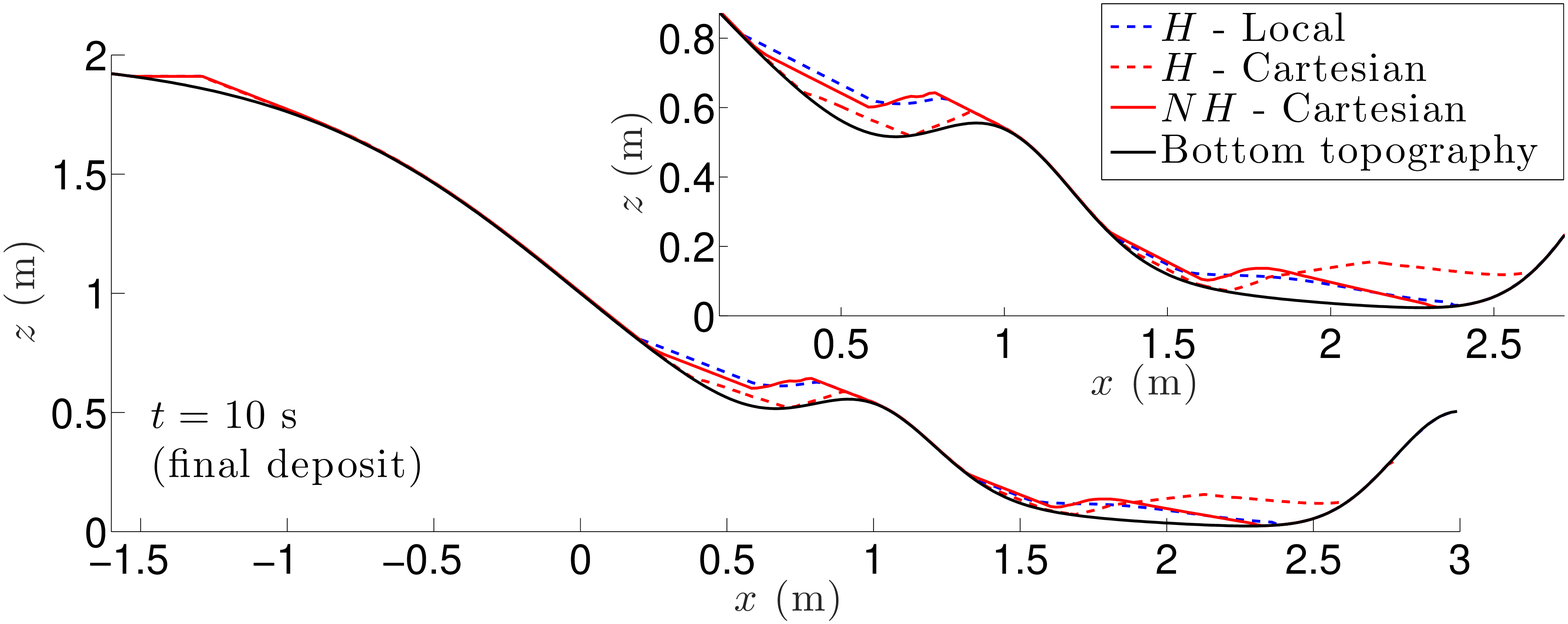}
	\end{center}
	\caption{\label{fig:test2} \it{Height of the flowing mass at different times. Dashed blue lines correspond to the local model with hydrostatic pressure, while red lines are the solutions of Cartesian models with hydrostatic (dashed) and non-hydrostatic (solid) pressures. Inset figures show zooms of the front position.}}
	\end{figure}

Figure \ref{fig:test2} shows the height at times $0.5,\, 1.5,\, 2.5$ s and the final deposit. We show the results of the hydrostatic local, the hydrostatic Cartesian and the non-hydrostatic Cartesian models. We see that the hydrostatic Cartesian model is the fastest one as observed previously. We also see that the results of the hydrostatic local model and the non-hydrostatic Cartesian models are close for $t> 2.3$ s. Actually, the final deposits computed with both models are similar even though the dynamics differ. Moreover, the solution of the hydrostatic Cartesian model widely differs from the other models.

	These results confirm that the hydrostatic local model and the non-hydrostatic Cartesian model produce similar deposits even though the dynamics is different, but not too different in this test.  Moreover, in view of the results, the non-hydrostatic Cartesian model proposed here is an improvement of the model introduced in \cite{denlinger:2004}, in the sense that our model computes the vertical acceleration while their model uses an  estimate of this acceleration by taking the average of the vertical velocity deduced from the free surface and bottom boundary condition. \\
	
	In the rest of the paper, we shall only use local models.
	
	\subsection{Comparison with experimental granular collapses}
	In this section we compare the results of the hydrostatic and the non-hydrostatic models with experimental data detailed in \cite{mangeney:2010}. In these experiments, we have a granular column of height $h_0=14$ cm and length $L = 20$ cm, which is initially confined in a tank. The gate is opened so that the material is released from rest and flows over an inclined plane with slope $\theta\ge0$. We consider here five different slopes, $\theta = 0^{\circ},9.78^{\circ}, 16^{\circ},19^{\circ}$ and $22^{\circ}$. The bed is made of the same particles glued on it.
	
	The computational domain is $[-0.2,3]$ m and the initial height is given by
	$$
	h_0(x) = \left\{\begin{array}{ll}
	0.14 &\mbox{if } x\leq 0;\\
	0 &\mbox{otherwise.}
	\end{array}\right.
	$$
	
	\begin{figure}
		\begin{center}
			\includegraphics[width=0.99\textwidth]{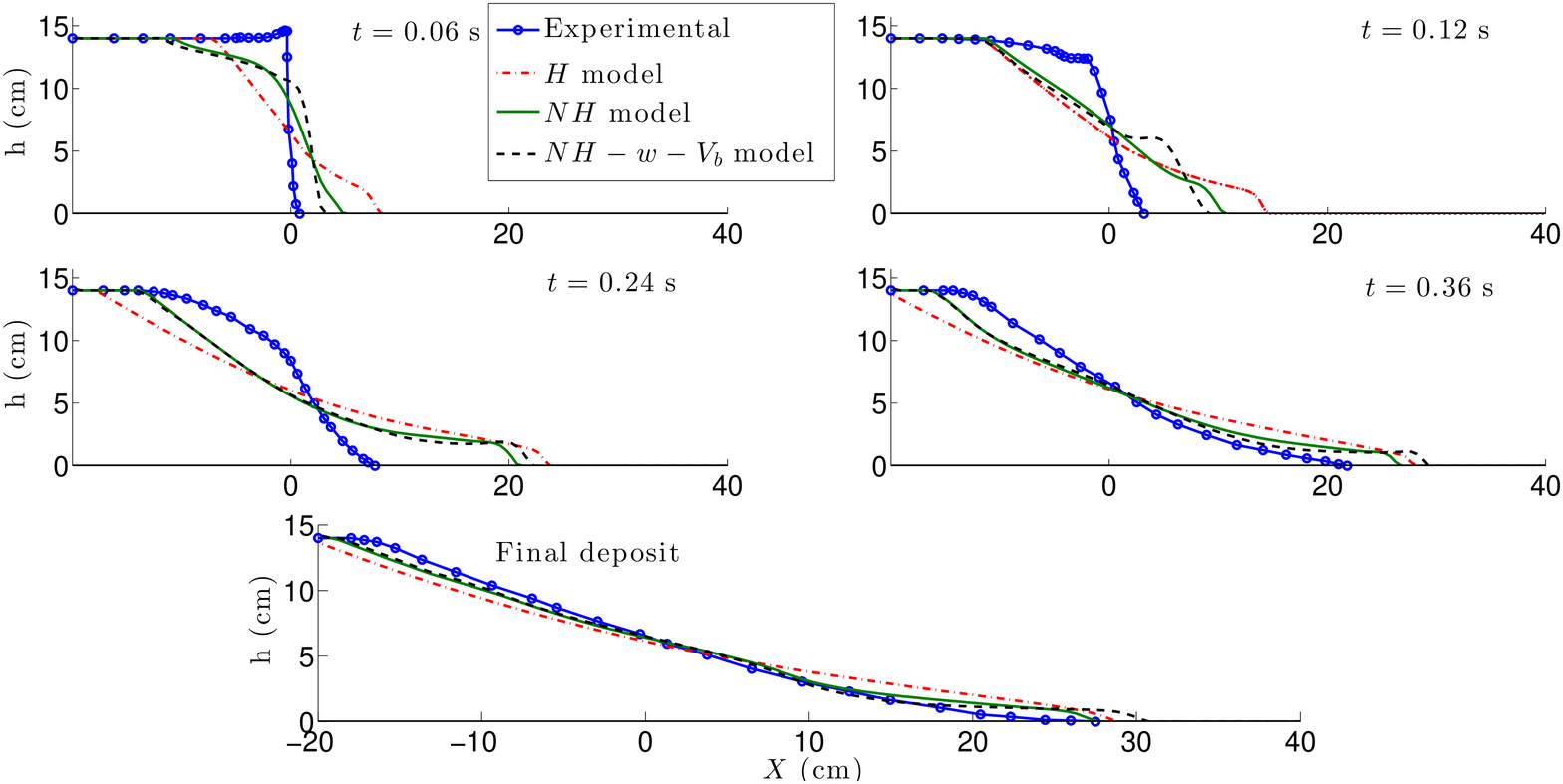}
		\end{center}
		\caption{\label{fig:0grados} \it{Time evolution of the granular mass with slope $\theta = 0^\circ$, for the laboratory experiments (solid-circle blue line), the hydrostatic model (dot-dashed red line), the non-hydrostatic model (solid green line) and the non-hydrostatic model with gate effect (dashed brown line).}}
	\end{figure}
	
	\begin{figure}
		\begin{center}
			\includegraphics[width=0.99\textwidth]{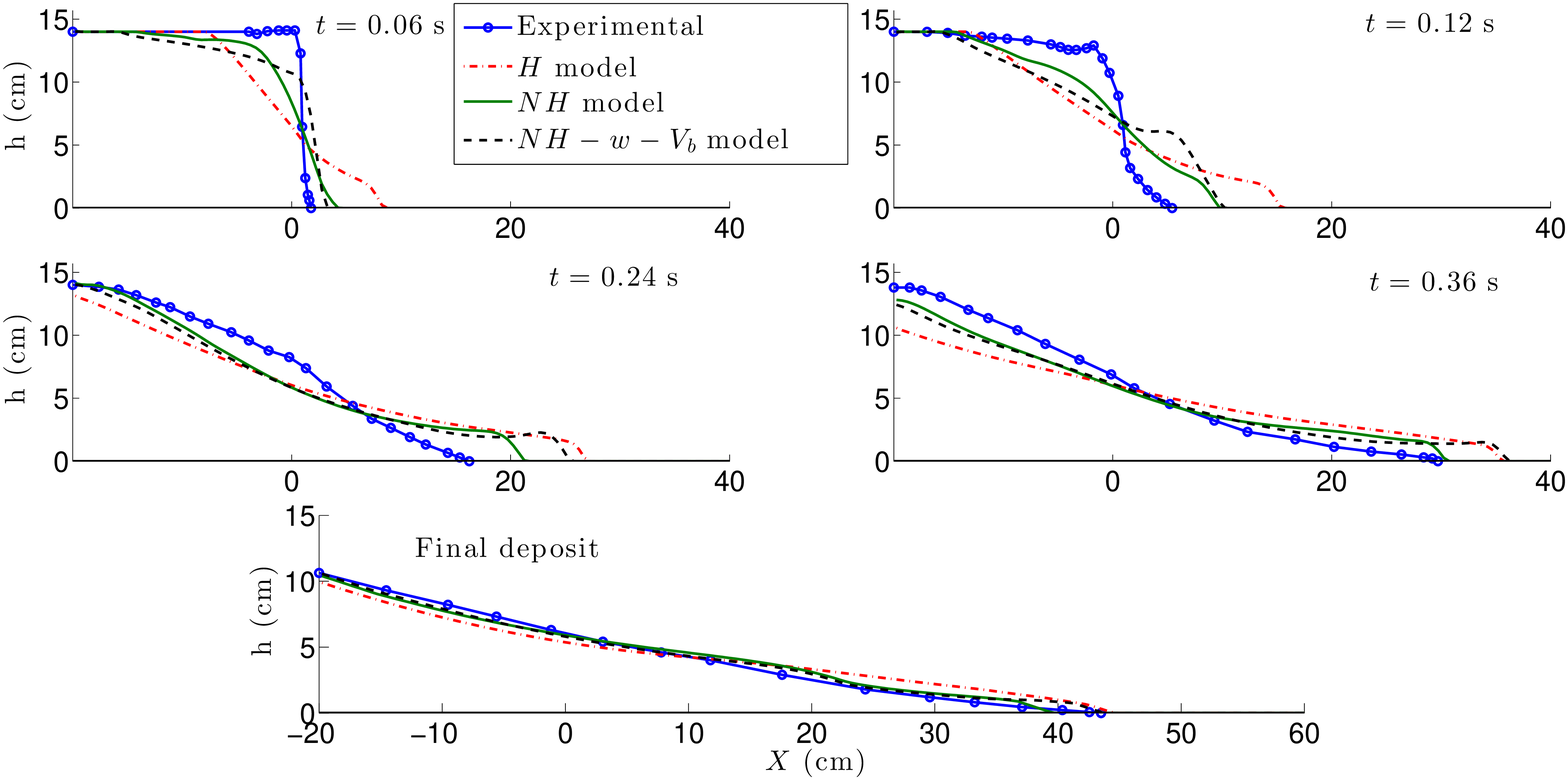}
		\end{center}
		\caption{\label{fig:978grados} \it{Time evolution of the granular mass over a plane with slope $\theta = 9.78^\circ$, for the laboratory experiments (solid-circle blue line), the hydrostatic model (dot-dashed red line), the non-hydrostatic model (solid green line) and the non-hydrostatic model with gate effect (dashed brown line).}}
	\end{figure}
	
	\begin{figure}
		\begin{center}
			\includegraphics[width=0.99\textwidth]{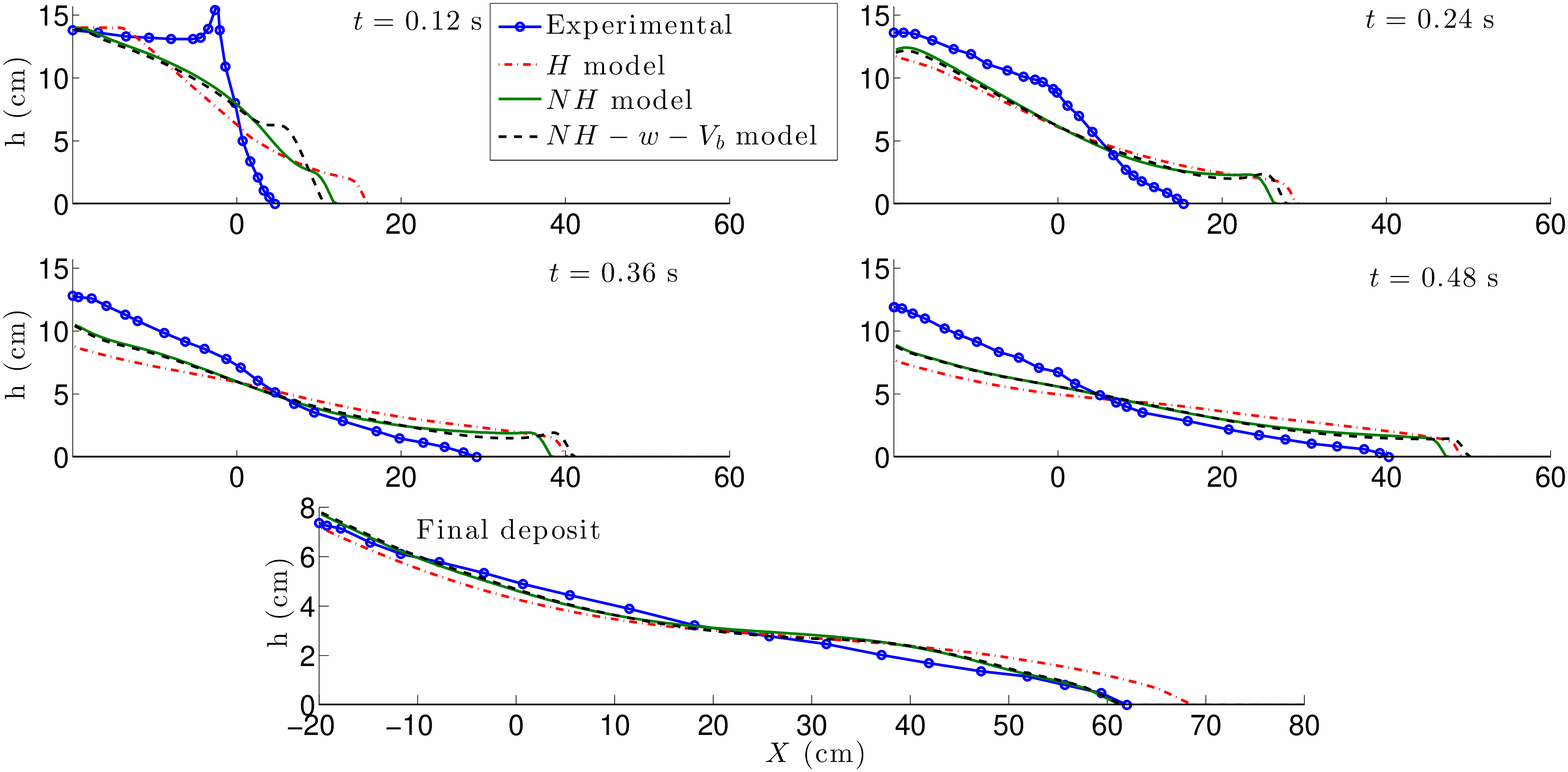}
		\end{center}
		\caption{\label{fig:16grados} \it{Time evolution of the granular mass over a plane with slope $\theta = 16^\circ$, for the laboratory experiments (solid-circle blue line), the hydrostatic model (dot-dashed red line), the non-hydrostatic model (solid green line) and the non-hydrostatic model with gate effect (dashed brown line).}}
	\end{figure}

	\begin{figure}
		\begin{center}
			\includegraphics[width=0.99\textwidth]{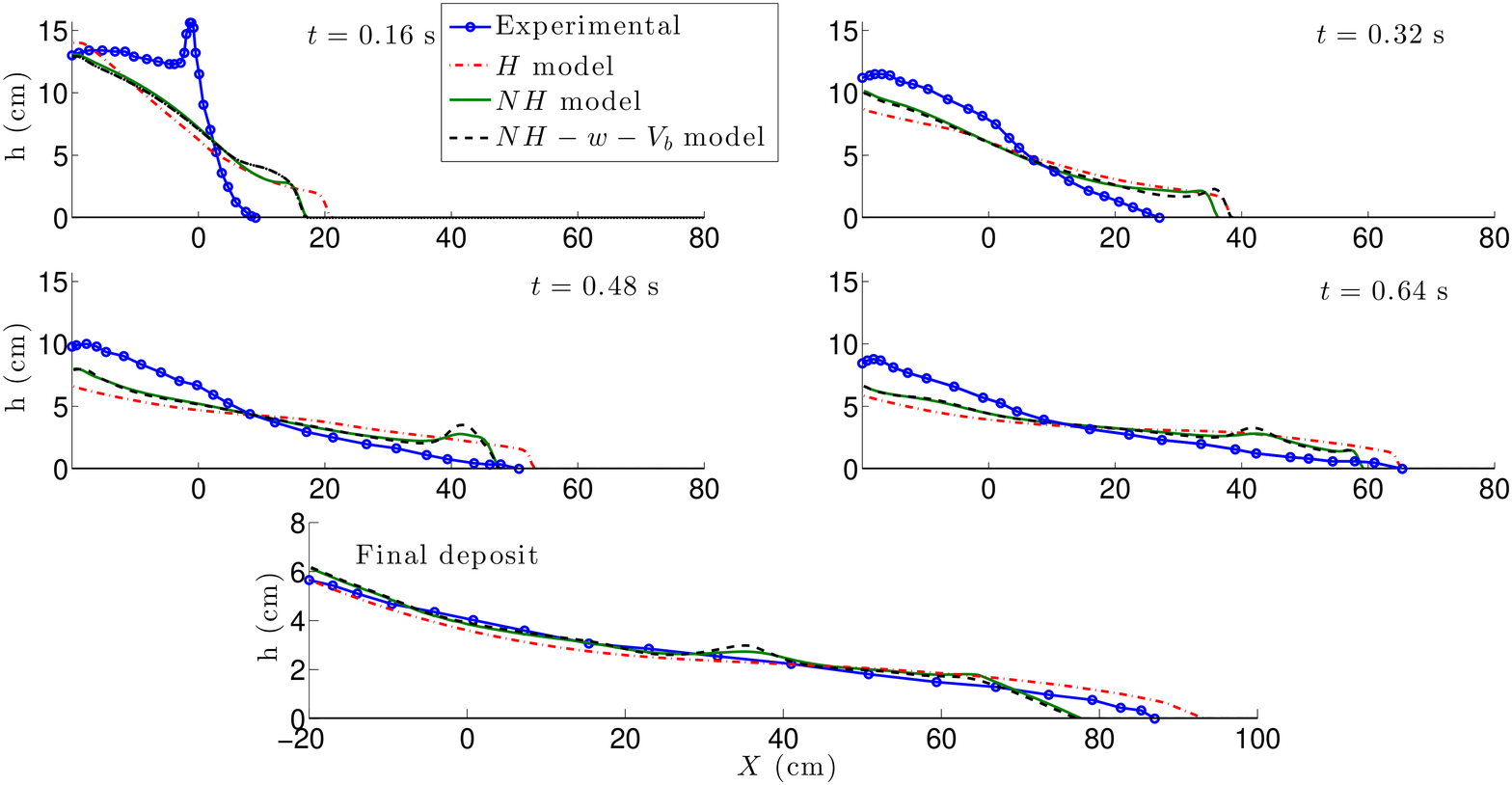}
		\end{center}
		\caption{\label{fig:19grados} \it{Time evolution of the granular mass over a plane with slope $\theta = 19^\circ$, for the laboratory experiments (solid-circle blue line), the hydrostatic model (dot-dashed red line), the non-hydrostatic model (solid green line) and the non-hydrostatic model with gate effect (dashed brown).}}
	\end{figure}
	
	\medskip
	
	\begin{figure}
		\begin{center}
			\includegraphics[width=0.99\textwidth]{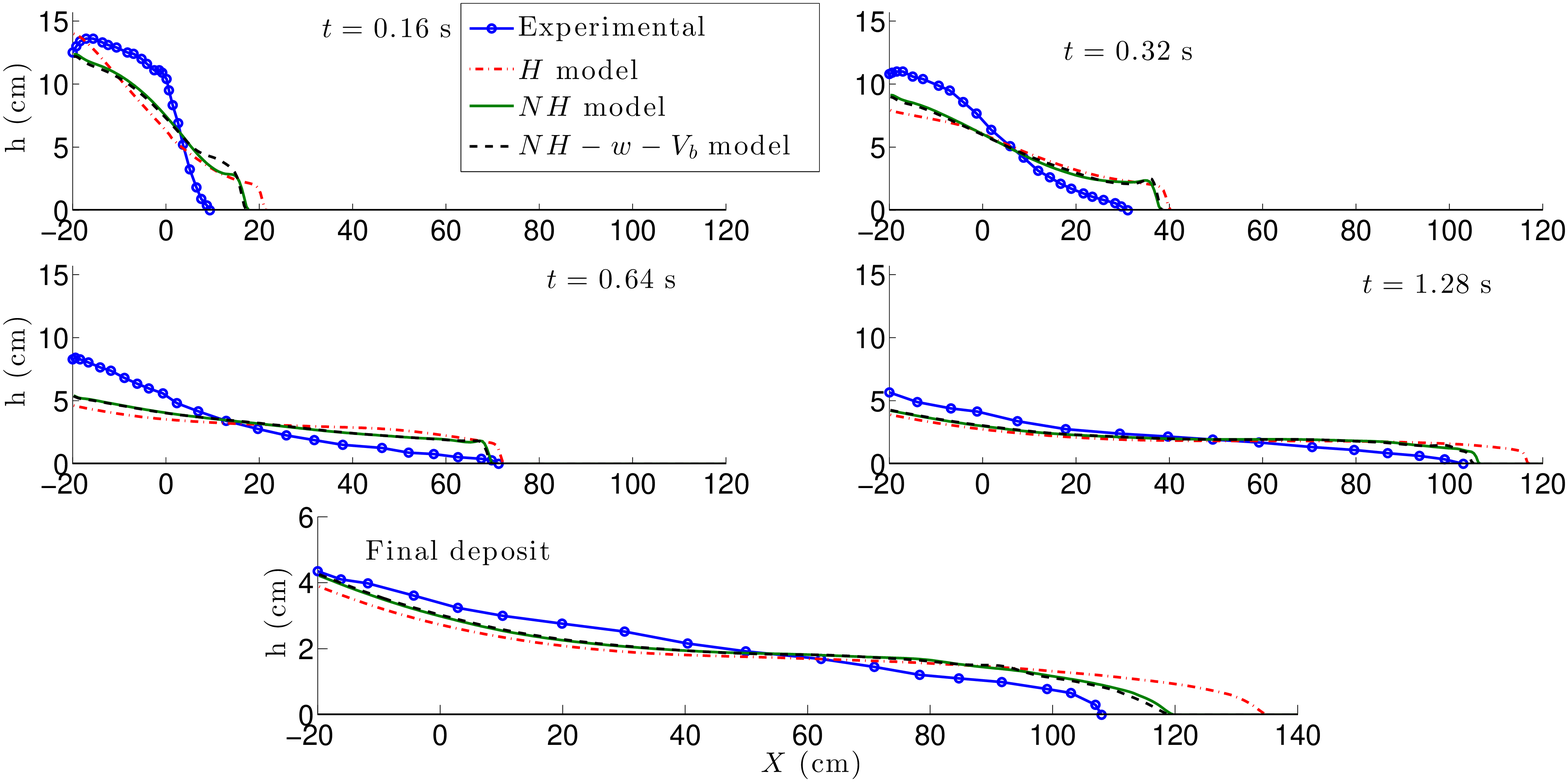}
		\end{center}
		\caption{\label{fig:22grados} \it{Time evolution of the granular mass over a plane with slope $\theta = 22^\circ$, for the laboratory experiments (solid-circle blue line), the hydrostatic model (dot-dashed red line), the non-hydrostatic model (solid green line) and the non-hydrostatic model with gate effect (dashed brown).}}
	\end{figure}
	
	\begin{figure}[!ht]
		\begin{center}
			\includegraphics[width=0.40\textwidth]{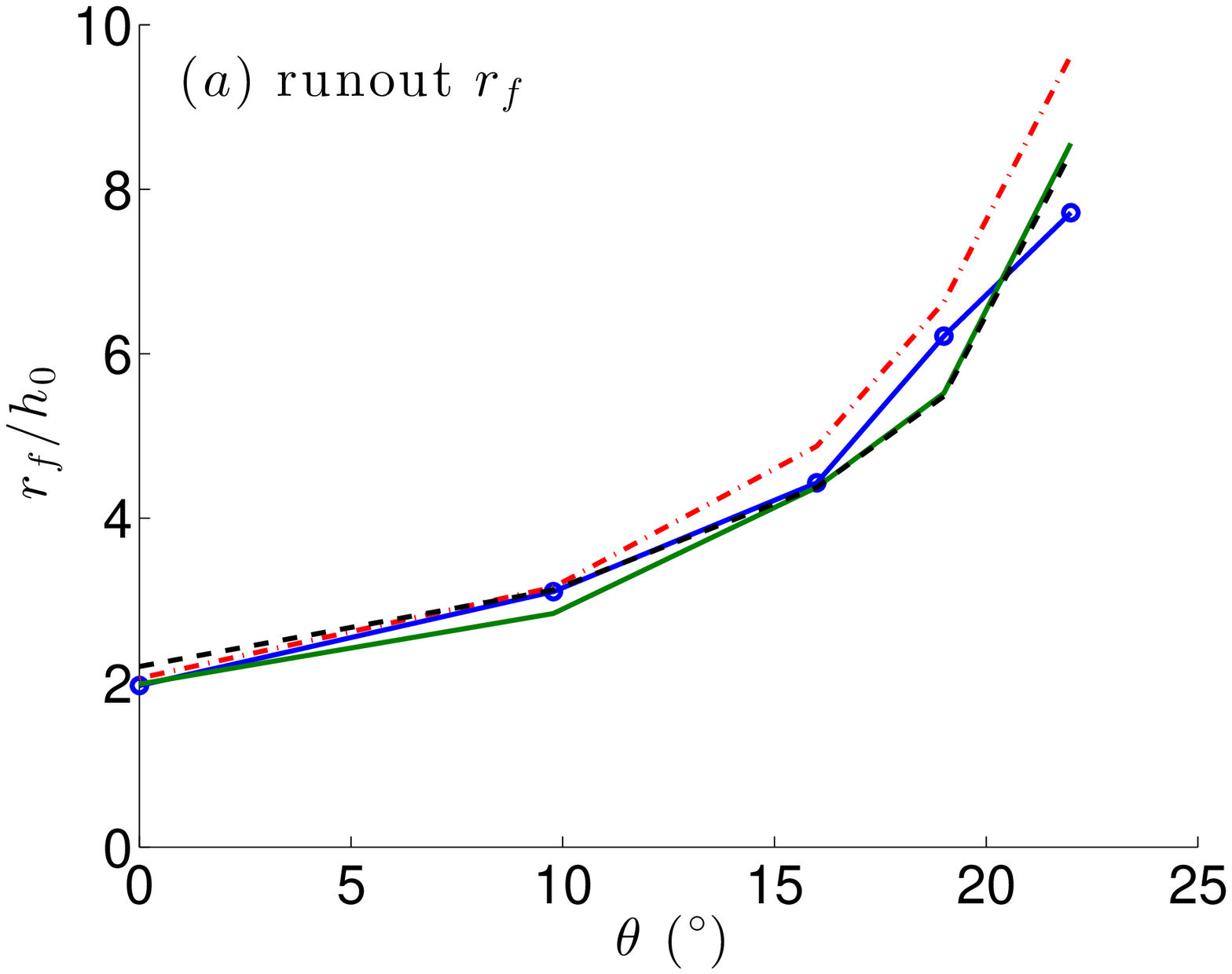}
			\includegraphics[width=0.40\textwidth]{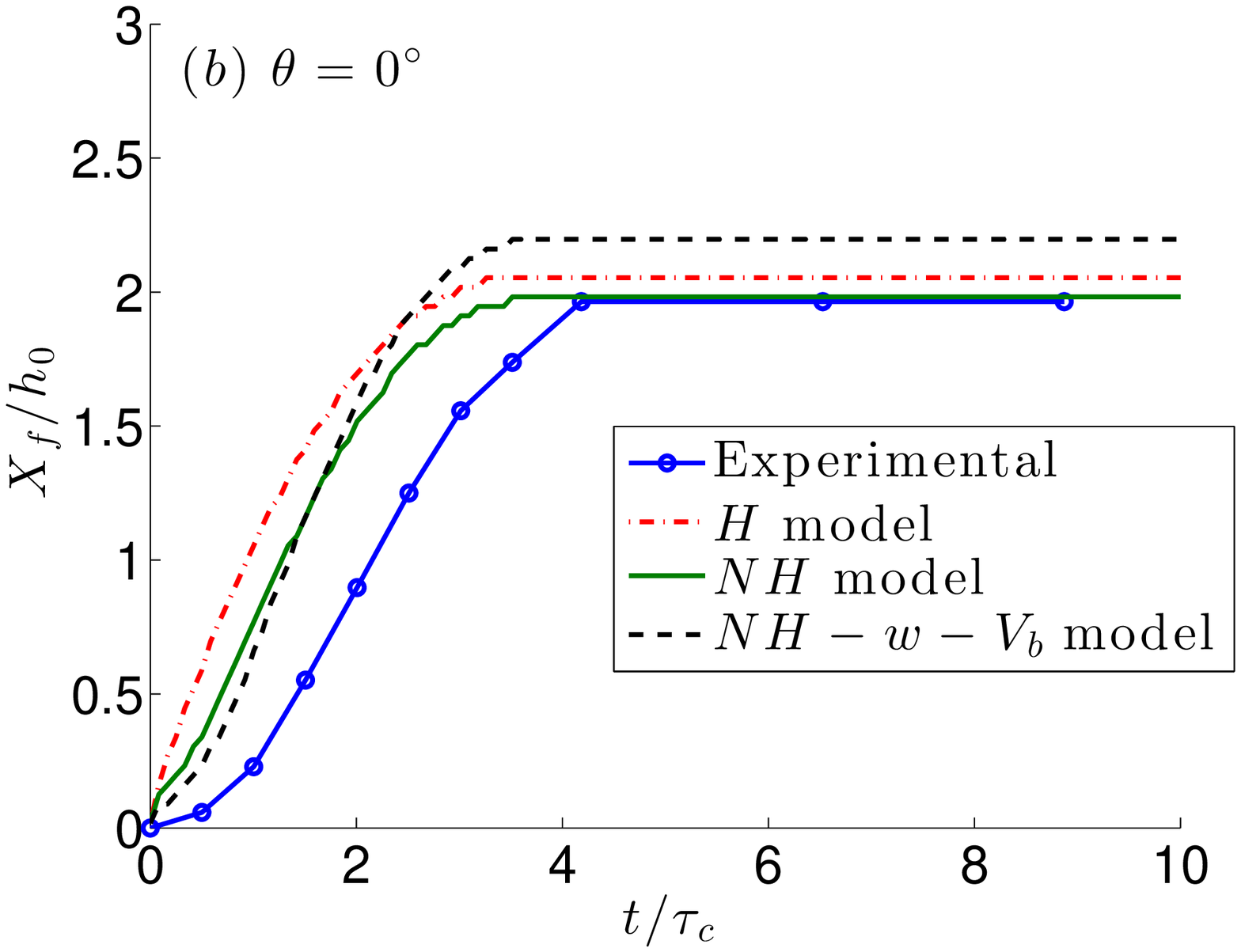}
			\includegraphics[width=0.40\textwidth]{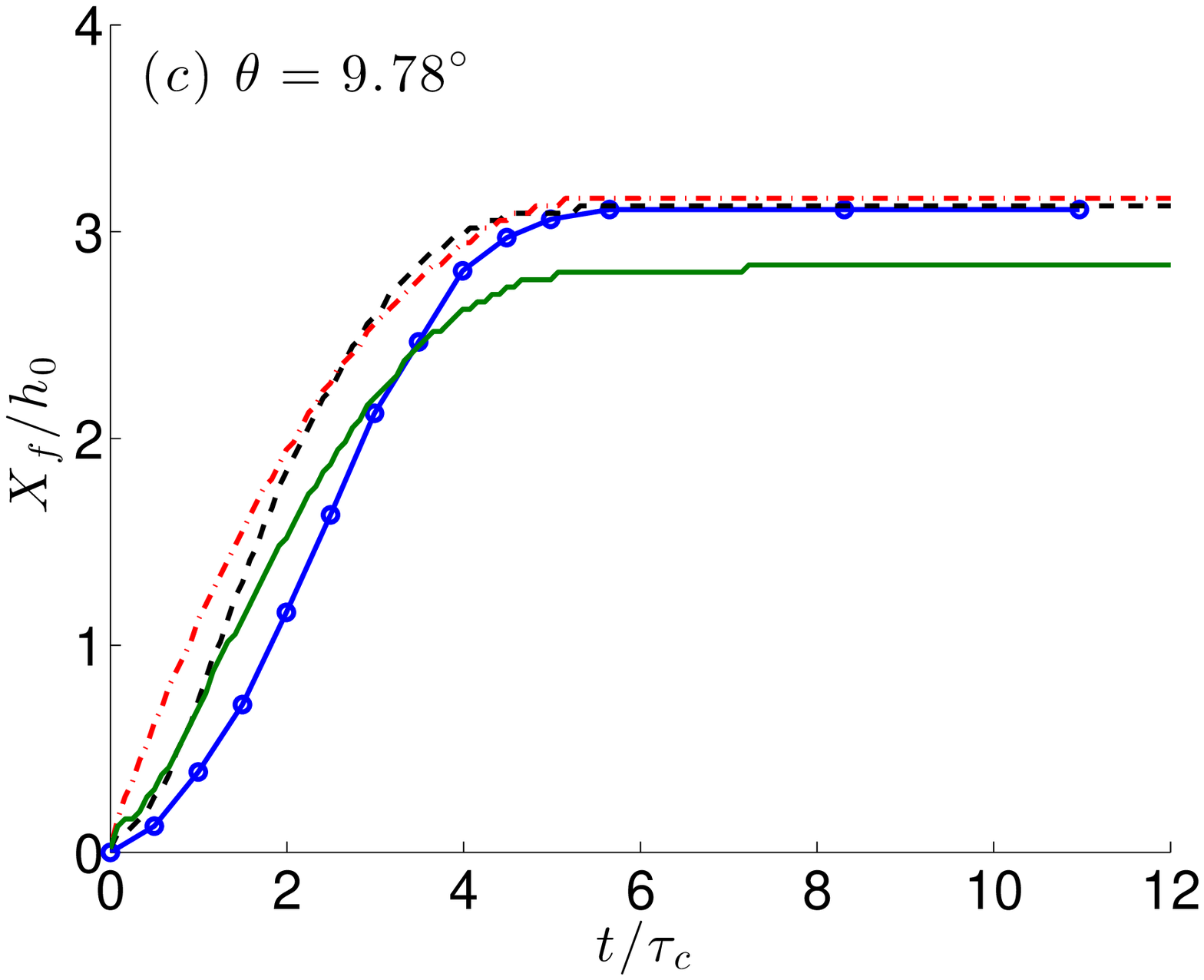}
			\includegraphics[width=0.40\textwidth]{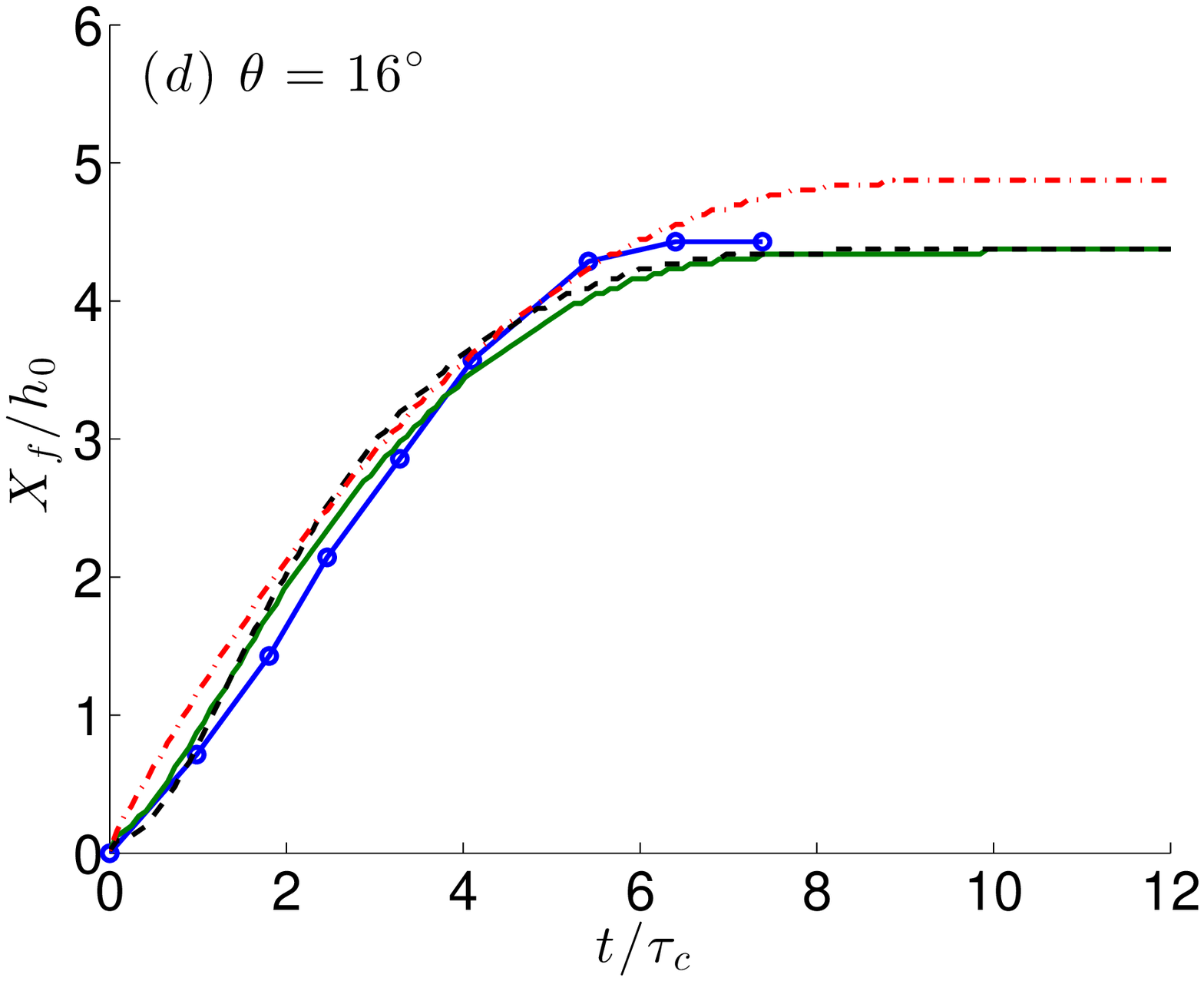}
			\includegraphics[width=0.40\textwidth]{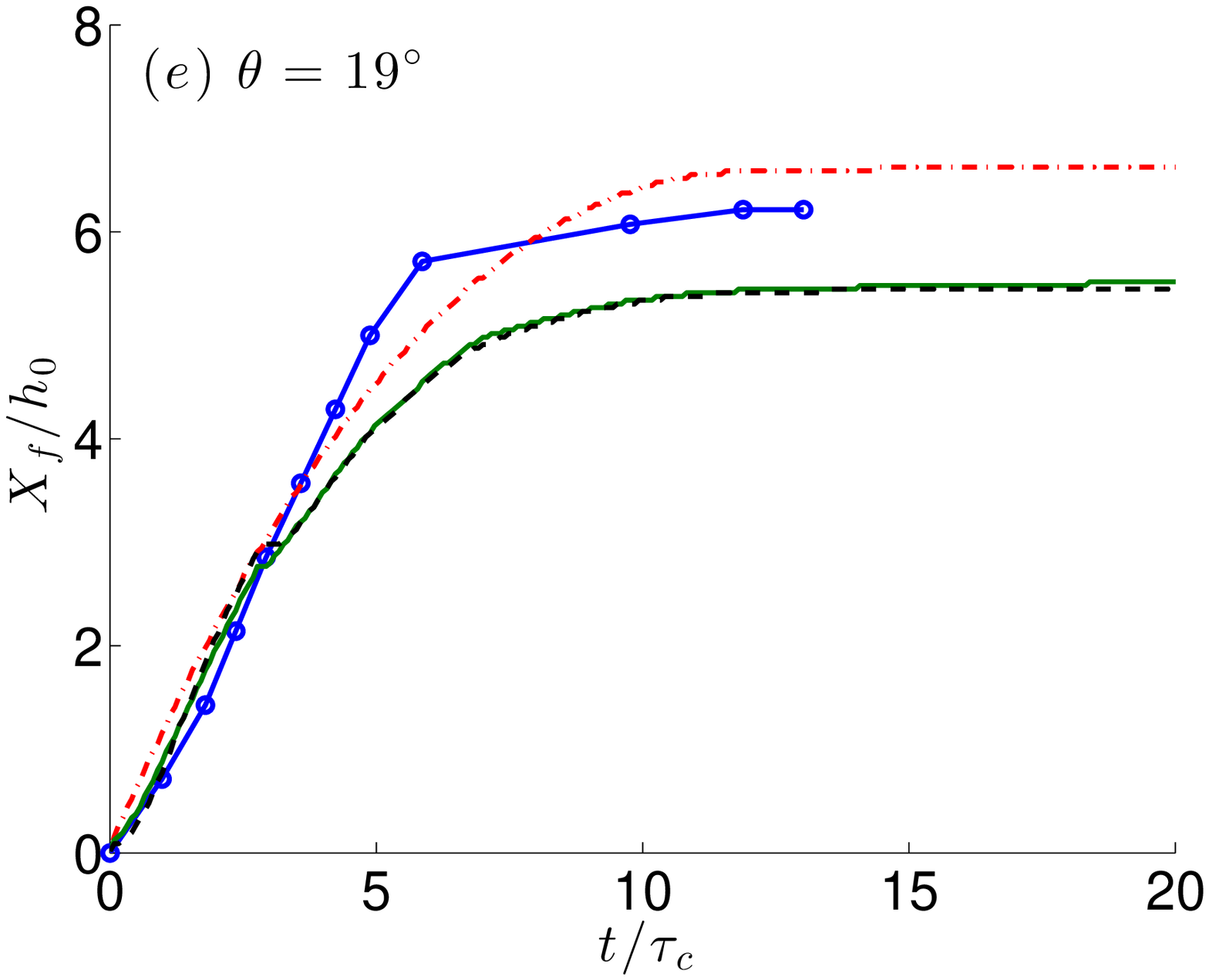}
			\includegraphics[width=0.40\textwidth]{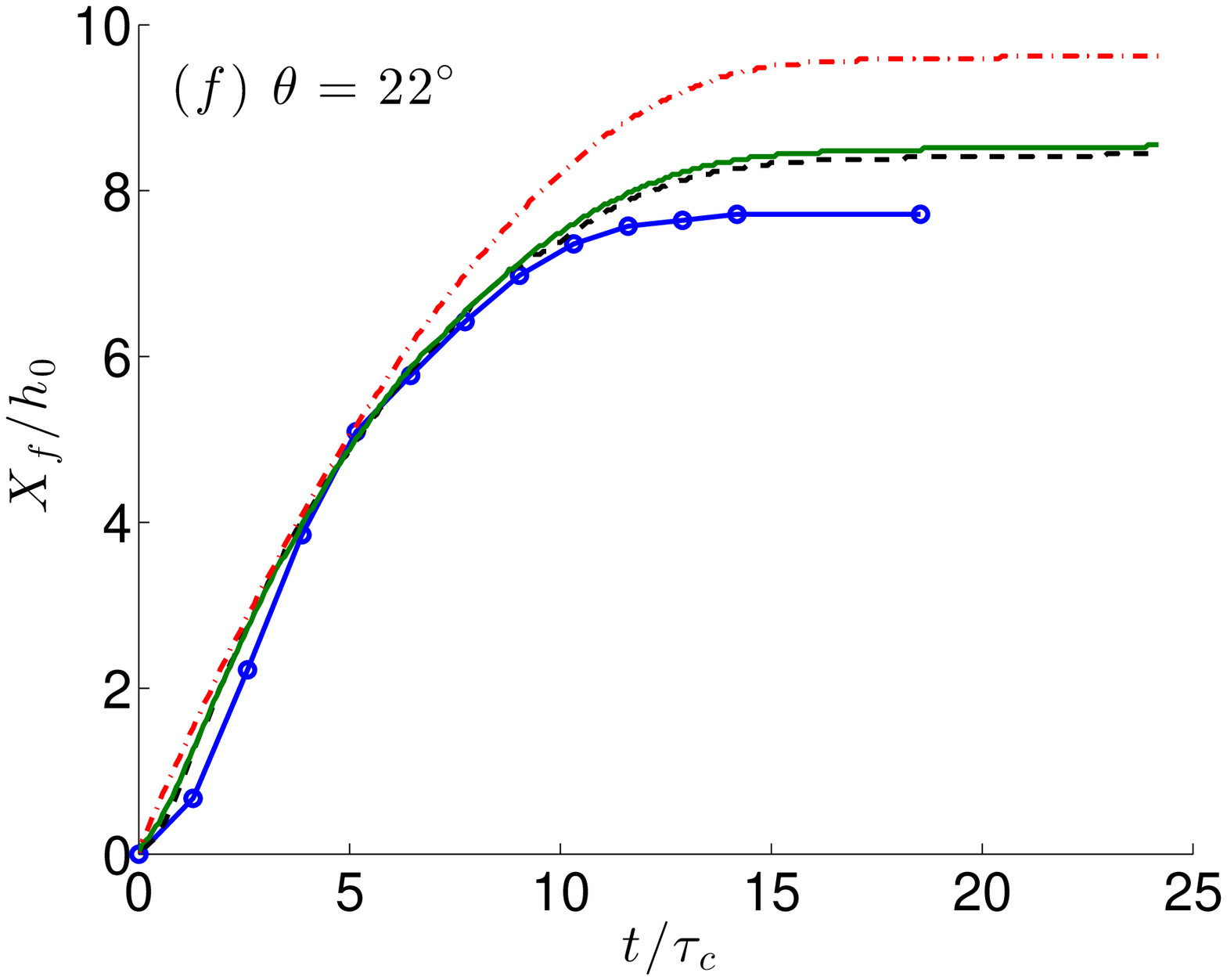}
		\end{center}
		\caption{\label{fig:frontposition} \it{(a) Normalized runout for all the slopes; (b)-(f) time evolution of the normalized position of the front computed with the hydrostatic model (dot-dashed red line), the non-hydrostatic model (solid green line), the non-hydrostatic model with the effect of the gate at initial times (dashed brown line), and experimental data (solid-circle blue lines). $h_0=0.14$ m and $\tau_c = \sqrt{h_0/(g\cos\theta)}$ s.}}
	\end{figure}
	\begin{figure}
		\begin{center}
			\includegraphics[width=0.4\textwidth]{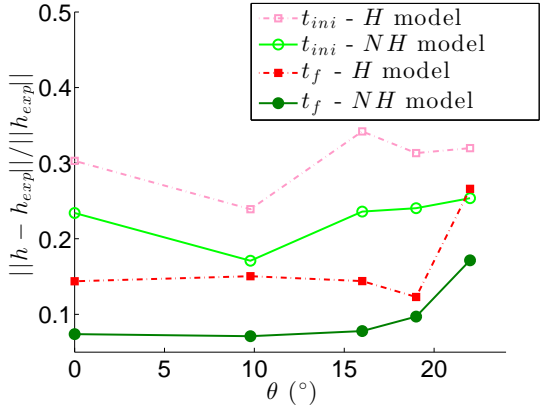}
		\end{center}
		\caption{\label{fig:error_h_theta} \it{Relative errors of the height along the domain computed with the hydrostatic (dot-dashed lines) and non-hydrostatic (solid lines) models. Filled-symbols lines correspond to the errors at final times $t_f$, while empty-symbols lines are the errors at time $t_{ini} = 0.24\, (0^\circ,9.78^\circ), 0.36 \,(16^\circ)$, $0.32\, (19^\circ,22^\circ)$ s}.}
	\end{figure}
	\begin{figure}[!ht]
		\begin{center}
			\includegraphics[width=0.49\textwidth]{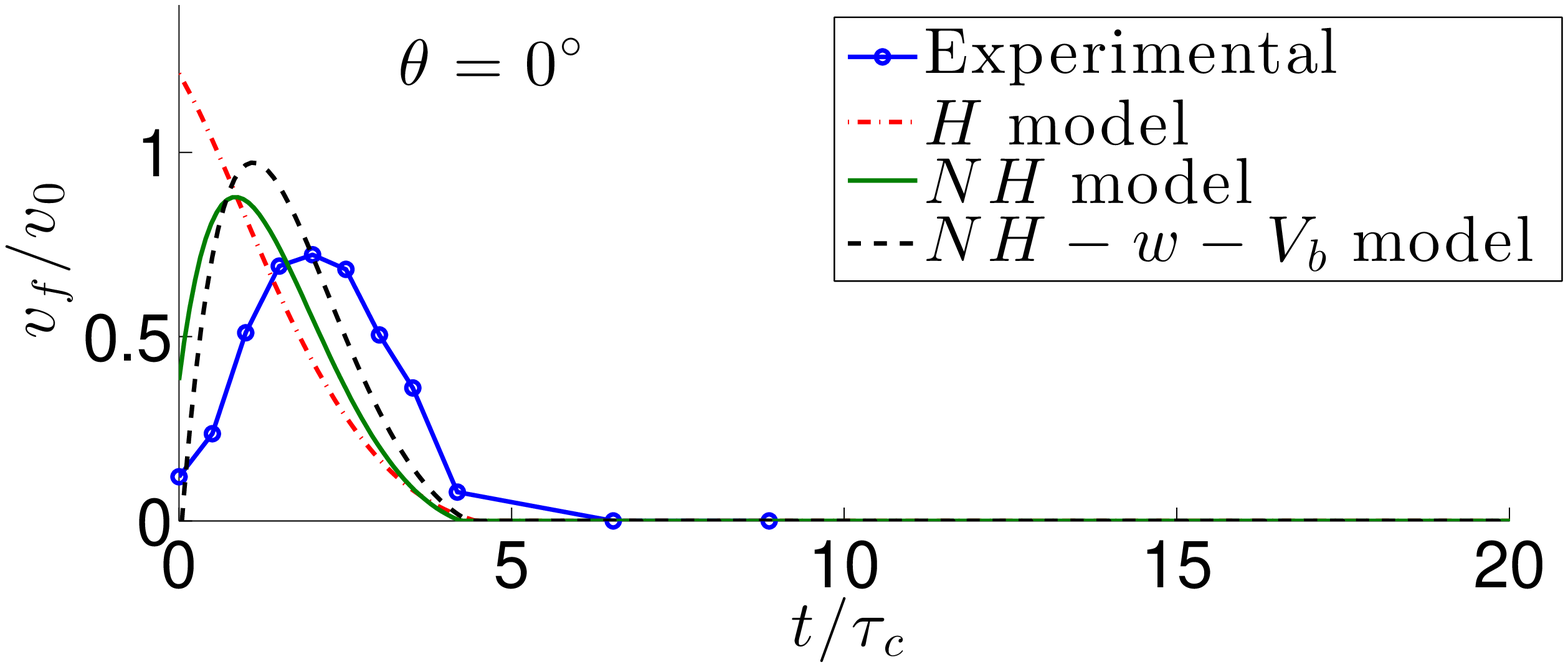}
			\includegraphics[width=0.49\textwidth]{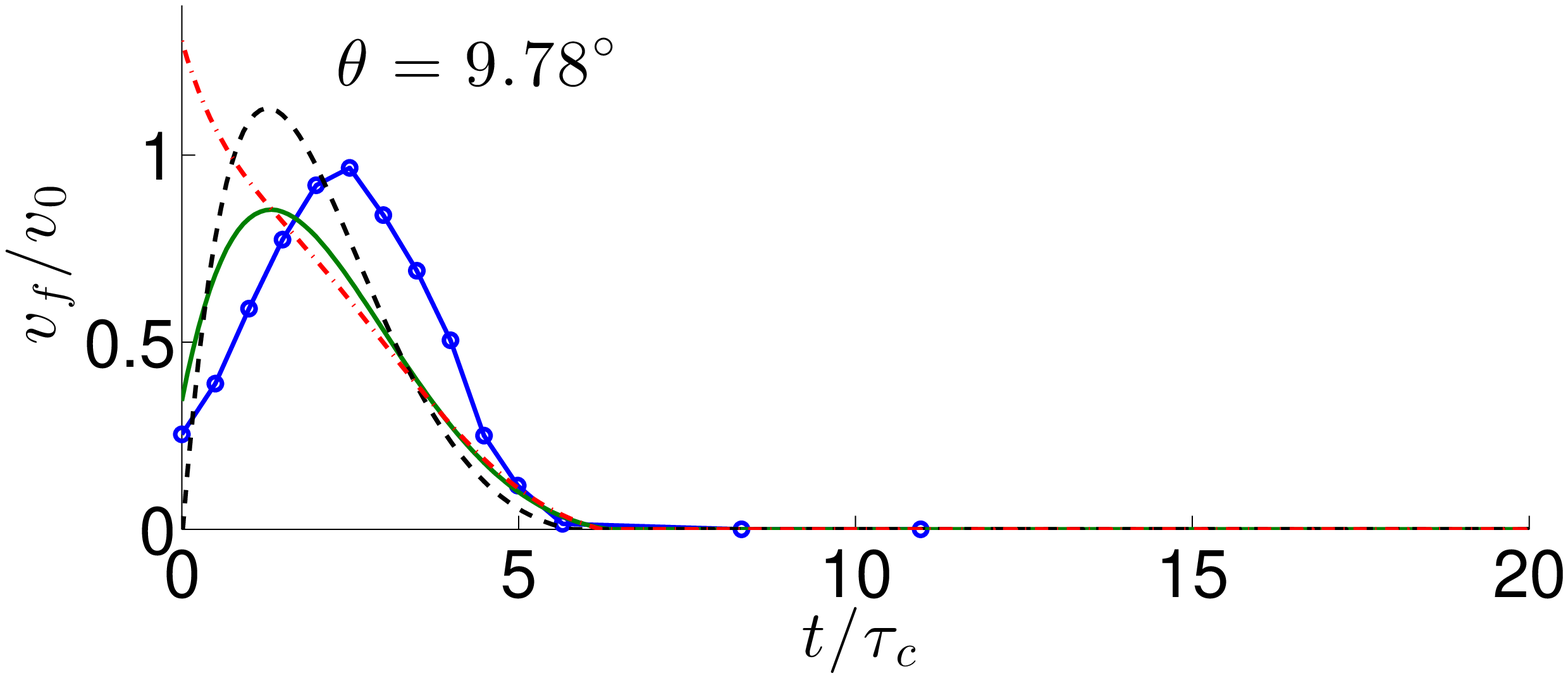}
			\includegraphics[width=0.49\textwidth]{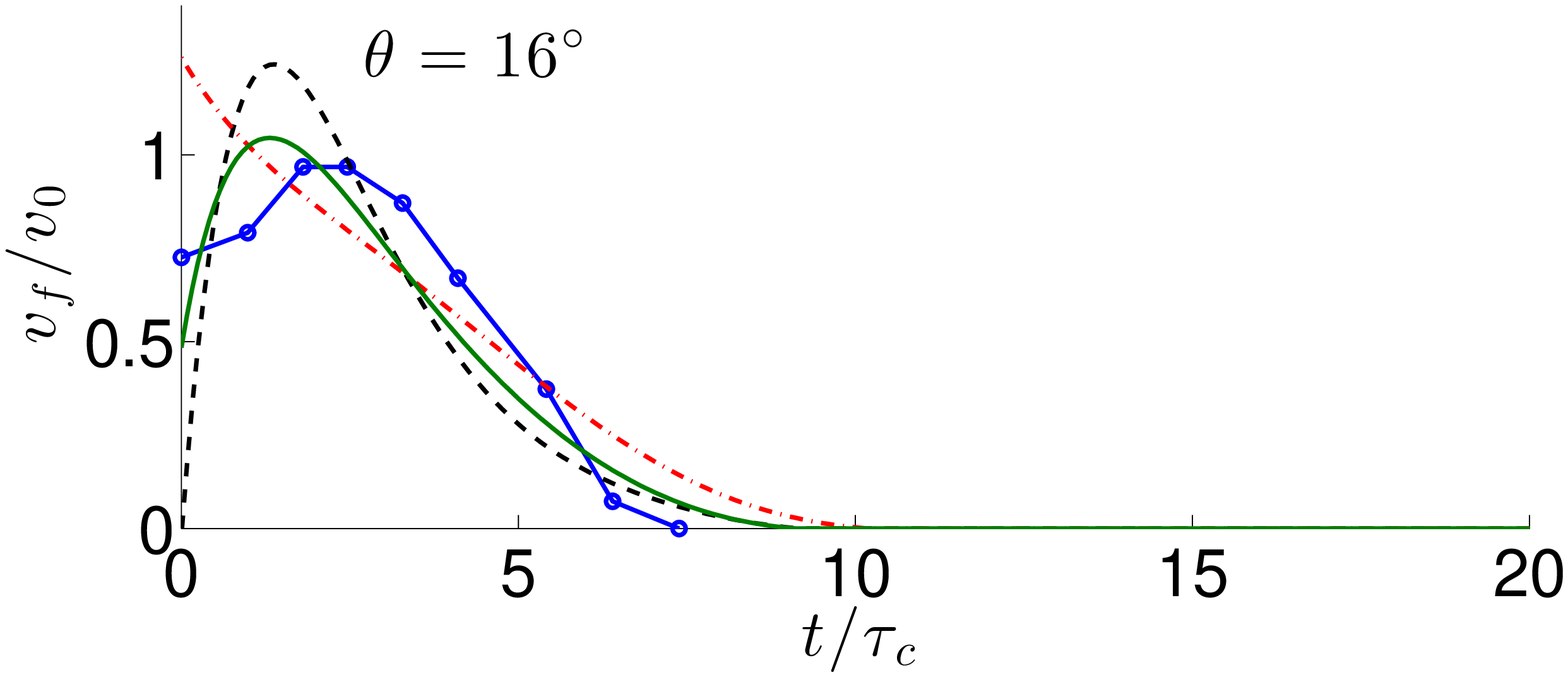}
			\includegraphics[width=0.49\textwidth]{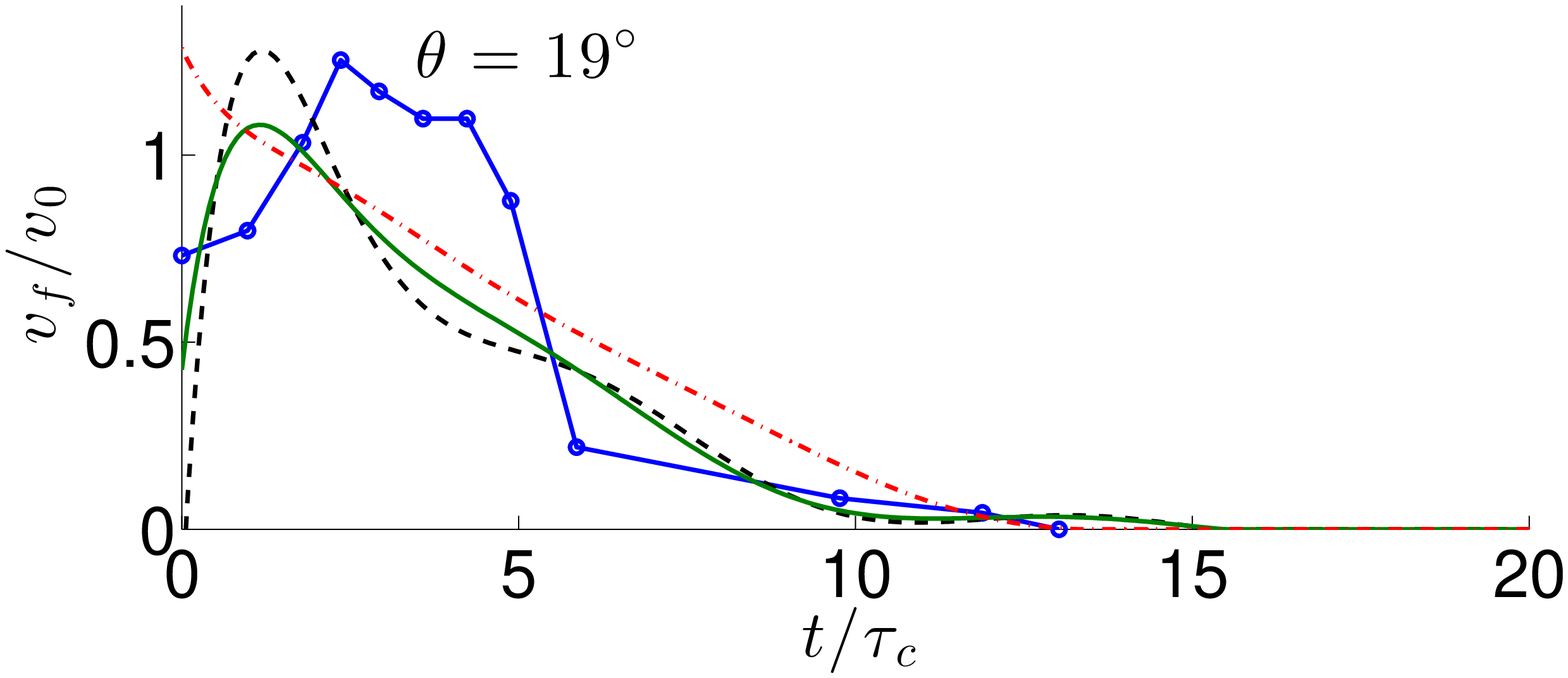}
			\includegraphics[width=0.49\textwidth]{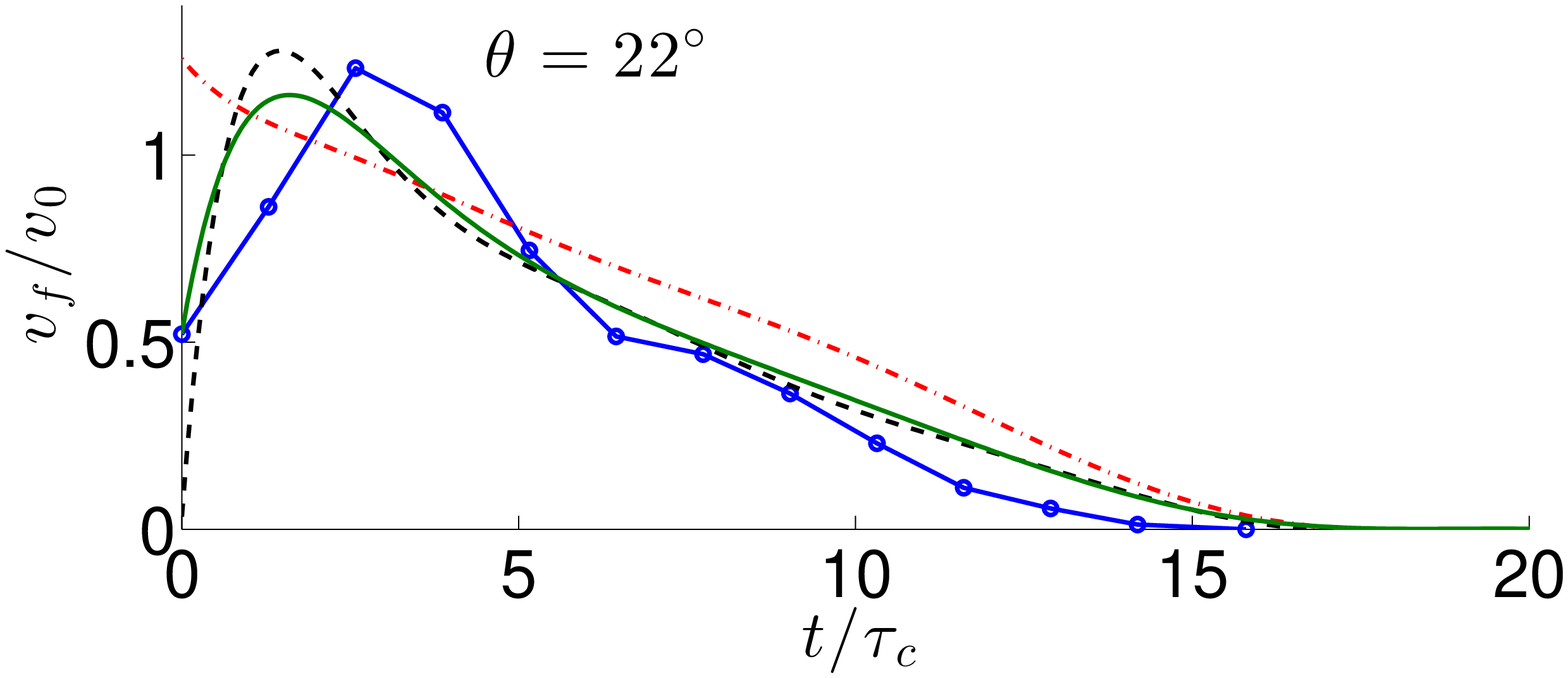}
		\end{center}
		\caption{\label{fig:frontvelocity} \it{Time evolution of the normalized velocity of the front computed with the hydrostatic model (dot-dashed red line), the non-hydrostatic model (solid green line), the non-hydrostatic model with the effect of the gate at initial times (dashed brown line), and experimental data (solid-circle blue lines). $h_0=0.14$ m, $v_0 = \sqrt{h_0 g\cos\theta}$ m/s and $\tau_c = \sqrt{h_0/(g\cos\theta)}$ s.}}
	\end{figure}
\begin{figure}[!ht]
	\begin{center}
		\includegraphics[width=0.45\textwidth]{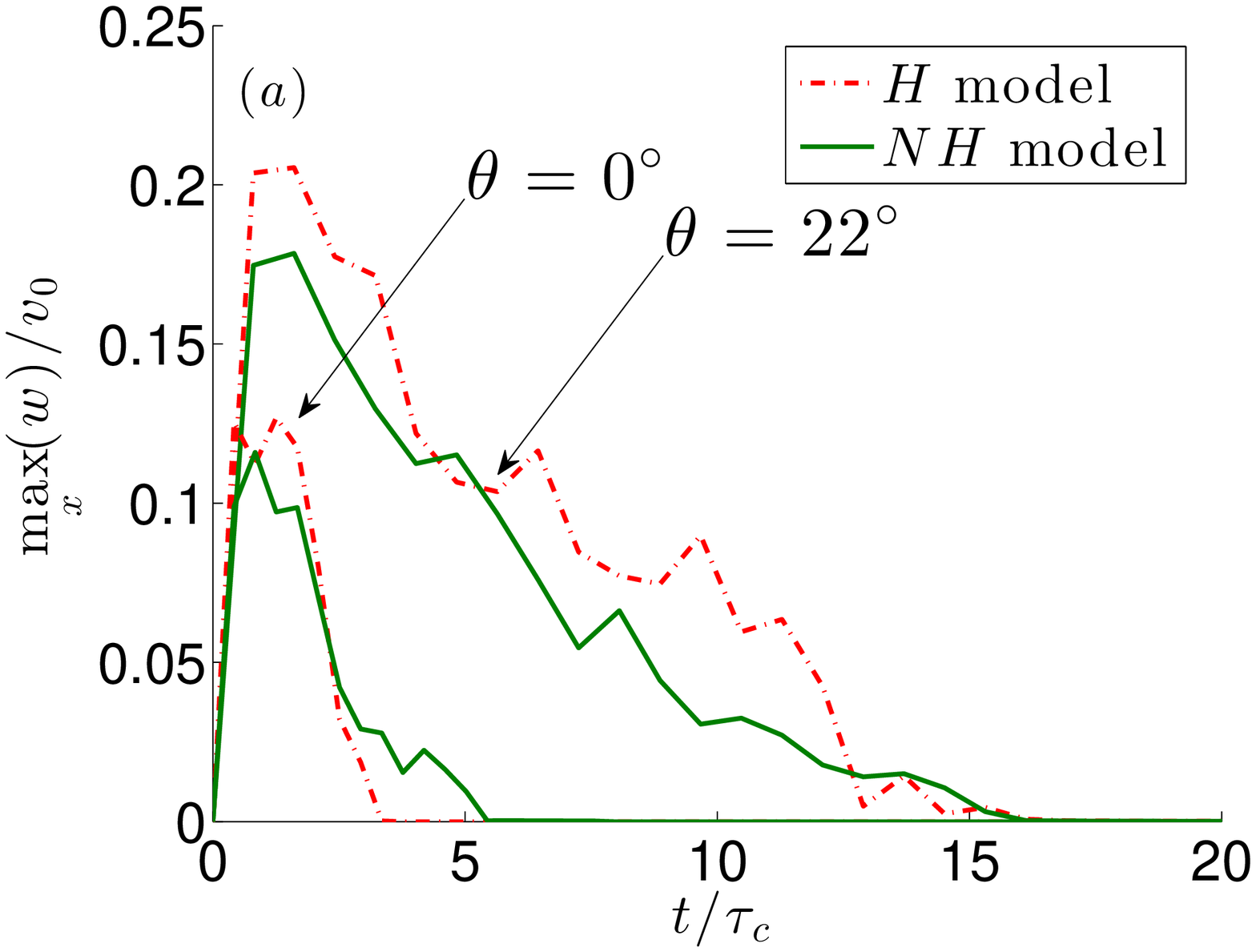}		
		\includegraphics[width=0.45\textwidth]{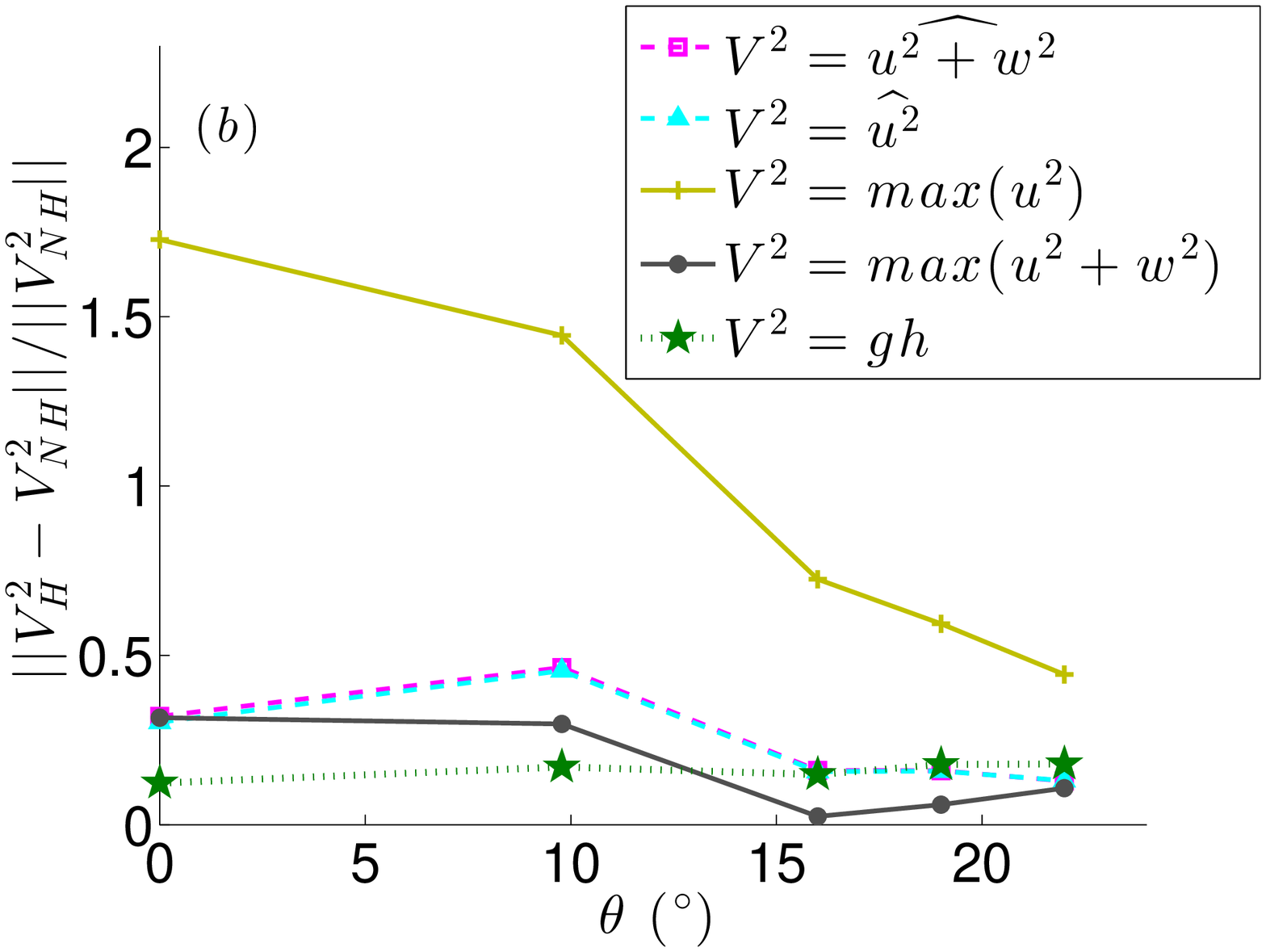}
	\end{center}
	\caption{\label{fig:errors_velocity} \it{(a) Time evolution of the maximum of the vertical velocity for $\theta = 0^\circ, 22^\circ$ computed with the hydrostatic (dot-dashed red lines) and the non-hydrostatic (solid green lines) model. $h_0=0.14$ m, $v_0 = \sqrt{h_0 g\cos\theta}$ m/s and $\tau_c = \sqrt{h_0/(g\cos\theta)}$ s. (b) Maximum on time of the differences of estimations of the energy between hydrostatic and non-hydrostatic models for all the slopes. $\max(\cdot) = \max_{t,x}(\cdot)$ and $\widehat{(\cdot)} = \max_{t}\left(\sum_{1\leq i\leq N} (\cdot)/N\right)$. }}
\end{figure}

Figures \ref{fig:0grados}, \ref{fig:978grados}, \ref{fig:16grados}, \ref{fig:19grados} and \ref{fig:22grados} show the thickness of the granular mass at different times, and the final deposit obtained with the hydrostatic and the non-hydrostatic model for a slope $\theta = 0^\circ, 9.78^\circ, 16^\circ, 19^\circ$ and $22^\circ$. We shall in particular focus on the solutions at short times. The reason is that when the gate is opened and the material starts to flow, the non-hydrostatic effects are strong because the mass is not shallow with strong gradients of the free surface. The non-hydrostatic effects decrease as the mass spreads and gets closer to a shallow layer. One of the known consequence of these non-hydrostatic effects is that the granular mass does not start moving as fast as when using models based on the hydrostatic assumption (see e.g. \cite{mangeney:2005}, \cite{fernandezNieto:2016}).

We indeed observe that the solution of the non-hydrostatic model is slower and more accurate than the solution obtained with the hydrostatic model in all the studied configurations, as detailed below. The final deposits obtained with the non-hydrostatic model are in good agreement with the experiments, in particular for $\theta = 9.78^\circ, 16^\circ$ (see figures \ref{fig:978grados}, \ref{fig:16grados}).

We see this behavior even more clearly in Figure \ref{fig:frontposition}, where the position of the front at different times is showed. It is observed that the front position computed with the non-hydrostatic model is  closer to the one obtained in the laboratory experiments up to a certain time, even though at the final instants the front position simulated with the NH model may be less accurate than with the H model as also observed on Figure \ref{fig:frontposition}a, showing the final runout $r_f$ as function of the slope. Indeed in figures \ref{fig:frontposition}a,\ref{fig:frontposition}c,\ref{fig:frontposition}e the experimental runouts for $\theta = 9.78^\circ$ and $\theta=19^\circ$ are closer to the hydrostatic model than to the non-hydrostatic one.  
Nevertheless, for $\theta = 9.78^\circ$, looking at the final deposit in Figure \ref{fig:978grados}, we see that the experimental front has a very small thickness. When shifting the front position up slope, where the mass thickness is not too small, the runout distance gets closer to the NH results than to the H simulations. For $\theta = 19^\circ$, even though the runout distance is underestimated with the NH model, the whole granular thickness is better approximated with the NH model (see Figure \ref{fig:19grados}), as explained below and as represented in Figure \ref{fig:error_h_theta}.

In order to quantify how accurately the models reproduce laboratory experiments when including non-hydrostatic terms, we represent in Figure \ref{fig:error_h_theta} the relative error on the mass thickness between the simulation and the experiments averaged over all the domain at a given time. This error is computed at a given short time $t_{ini}$ (chosen as a time for which the flow is initialized and we have experimental data) and at the final time $t_f$. We see that the error obtained with the non-hydrostatic model is smaller than the one obtained with the hydrostatic model for all slopes and for both the short and the final times. In particular, for $\theta = 9.78^\circ$ at final time, the error computed with the non-hydrostatic model is $7\%$ approximately, whereas this error is greater than $15\%$ with the hydrostatic model. Figure \ref{fig:error_h_theta} also shows that the error is smaller at the final time than at the short time. One of the source of the error is related to the depth-averaged process as shown in \cite{fernandezNieto:2016,fernandezNieto:2018}. In particular the rounded shape of the front obtained in the simulations disappears when multi-layer models are used, i. e. when no depth-averaging is performed (compare e. g. Figure \ref{fig:22grados} of the present paper to Figure 14 of \cite{fernandezNieto:2016}). Note that the error seems to increase with increasing slope. This may be due to wall effects that are more and more important as the slope angle increases as shown in \cite{martin:2017}.

In the figures, we see a small peak in the experiments at very short times. This results from the gate opening when releasing the granular material. The gate removal induces a vertical velocity to the material that is located near the front, in contact with the gate. This cannot be reproduced by hydrostatic models, but using non-hydrostatic models we can impose the initial vertical velocity induced by the gate removal.
Therefore, an advantage of the non-hydrostatic model is that we can impose a vertical velocity at initial time, in order to reproduce the vertical velocity induced by the gate opening.

To do that, we consider a initial vertical velocity defined by
\begin{equation}\label{eq:vert_vel_test}
W_b(x) = \left\{\begin{array}{ll}
V_{b} &\mbox{if } -0.025<x<0;\\
0 &\mbox{otherwise,}
\end{array}\right.
\end{equation}
where $V_b$, is the estimated velocity at which the gate is removed. In the experiment it is estimated that $V_b=2.3$ $m/s$ (see \cite{ionescu:2015}). Remark that in \cite{ionescu:2015}, the gate is imposed as a moving wall boundary condition. This, together with the fact that they solved the full 2D Navier-Stokes system, allows them to obtain better results. However, the computational effort is much bigger. In order to show how the non-hydrostatic model could be used to study the effect of the gate, we also have performed these numerical tests using this initial vertical velocity.

Figures \ref{fig:0grados}, \ref{fig:978grados}, \ref{fig:16grados}, \ref{fig:19grados}, \ref{fig:22grados} and \ref{fig:frontposition} show the results obtained when imposing this initial vertical velocity at the front position where the gate is located. We see that these results obtained with ($w_0(x)\neq 0$) are similar to the non-hydrostatic model starting from rest at large times, whereas they differ for short times. We also see in Figure \ref{fig:frontposition} that the evolution of the front position improves for short times when including the vertical velocity mimicking the gate removal, while this has almost no effect at final times. Moreover, we see that the influence of the gate is stronger for small slopes ($\theta=0^\circ,9.78^\circ$) than for larger slopes ($\theta=16^\circ,19^\circ,22^\circ$).
The results obtained here differ from those of \cite{ionescu:2015}, which used the full 2D Navier-Stokes equations (i. e. fully non-hydrostatic pressure) and described the gate as a moving boundary without friction. Indeed, in their simulation, gate effects make the front to propagate more slowly at the beginning as observed here, but the runout distance is independent of the gate, opposed to what is obtained here for $\theta=0^\circ$ and $\theta=9.78^\circ$ (see figures \ref{fig:0grados} and \ref{fig:978grados}, respectively).

Figure \ref{fig:frontvelocity} shows the velocity of the front for all the slopes. In the experiments we see that the velocity grows up at the beginning, and it decreases after an intermediate time, describing thus a parabolic profile. This behavior is reproduced with the non-hydrostatic model. On the contrary, the front velocity computed with the hydrostatic model starts from its maximum value and then decreases. This is an important improvement of NH models. Indeed, despite of being a simple model which neglects the first order contribution of the non-hydrostatic pressure \eqref{eq:pnh_antesdeintegrar_q1}, the shape of the front velocity is much better reproduced than with hydrostatic models.

In Figure \ref{fig:frontvelocity} we also see that the front velocity is smaller during the first instants for the model including an initial vertical velocity $w_0\neq 0$. Next, its growth is faster and the maximum velocity of the front is larger than the one computed with $w_{0}=0$. Interestingly, the maximum velocity is reached at similar times for both models.

Figure \ref{fig:errors_velocity} shows the time evolution of the maximum vertical velocity, which is reached close to the front, for the smallest and biggest studied slopes $\theta = 0^\circ, 22^\circ$. This velocity, when computed with the hydrostatic model, is bigger than with the non-hydrostatic model. This figure also shows the differences of estimated potential energy ($gh$) and kinetic energy ($V^2$) between the hydrostatic and the non-hydrostatic model, for all the slopes. In order to approximate the kinetic energy ($V^2$) we use the downslope velocity $u$ or the velocity vector $(u,w)$. For each case, we take the maximum on time, and the maximum or the average on space of $V^2$. Next, we compute the (relative) difference between the values computed with the hydrostatic and the non-hydrostatic model for a fixed slope. We see that, for small slopes, the difference between the models is significantly greater for the kinetic energy than for the potential one (which also represents the differences on the height), and these are of the same order of magnitude for all the slopes. However, the difference of the kinetic energy are greater for small values of the slope. We could conclude that the difference between both models is bigger for small slopes.

\section{Conclusions}\label{se:conclusions}

In this work a non-hydrostatic depth-averaged model for dry granular flows has been proposed. The model considers a friction term based on the $\mu(I)$ rheology, where the friction coefficient depends on both the pressure (hydrostatic and non-hydrostatic) and the velocity. For the sake of simplicity, we assume that  the non-hydrostatic pressure  has a linear profile. Otherwise, for other profiles, the system would have extra unknowns and equations resulting in more complexity from the computational point of view (see e.g. \cite{fernandezNieto:2018a}).

The  proposed model  notably improves the results of hydrostatic models, in particular when comparing our results with dam break laboratory experiments. The model can be seen as a correction of classical Savage-Hutter type models with a $\mu(I)$ friction law. Its numerical discretization can also be adapted for any existing hydrostatic code by adding two additional steps to the numerical scheme.

In the numerical tests, we have analyzed the influence of the coordinate system (Cartesian or local) for the hydrostatic and non-hydrostatic model. The non-hydrostatic models (both Cartesian and local) predict a slower motion of the granular front at the beginning. However, the front positions computed with both Cartesian models are longer after some time, as expected. It is due to the fact that Cartesian model use the horizontal velocity instead of the velocity tangent to the topography. The biggest differences between the NH local model and the H local, NH Cartesian, and H Cartesian models are found for short times (see Figure \ref{fig:local_cart_error_runout}b). Namely, the maximum of this difference is around 320$\%$ for the H Cartesian and 120$\%$ for the NH Cartesian models at time $t= 0.03$ s, whereas it is 46$\%$ for the H local model at $t=0.06$ s.

In addition, the deposits obtained with the local hydrostatic model and the Cartesian non-hydrostatic models are similar even though the dynamics differs. These results partly support the assumption made by \cite{denlinger:2004} where a hydrostatic Cartesian model with a correction of the pressure based on an approximation of the vertical acceleration is proposed with the aim to avoid working in local curvilinear coordinates.
In that sense, our non-hydrostatic Cartesian model is an improvement of the one proposed in \cite{denlinger:2004}, since the vertical acceleration is computed and not estimated. This has been studied in test \ref{se:tests_HvsNH}, where a complex topography has been used, obtaining similar conclusions.
We have also observed that the non-hydrostatic models are less dependent on the coordinate system than the hydrostatic models, which is also an expected result.

Comparisons have been made with laboratory experiments, and also with a hydrostatic model (Savage-Hutter model with a $\mu(I)$ friction coefficient). The non-hydrostatic model improves the results of the hydrostatic one, in particular at short times, which is clear by looking at the time evolution of the position of the front. The shape of the flowing mass and of the deposit is also always closer to the experiments when using non-hydrostatic models. The importance of non-hydrostatic terms is higher for smaller slopes, as expected.  The approximation of the position of the front is also improved when using the non-hydrostatic model. Moreover, we have shown that the thickness distribution is always better reproduced by the NH model compared to the H model, both at short and final times. For example, as shown in Figure \ref{fig:error_h_theta}, for $\theta \leq 16^\circ$ ($\theta = 22^\circ$, respectively) the relative error between simulated and observed thickness distribution is around 15$\%$ for the H model while it is approximately 7$\%$ for the NH model (27$\%$ and 17$\%$, respectively).

By using this non-hydrostatic model we may impose a vertical velocity at initial time, which mimics the effect of the gate opening. The gate opening has a strong influence on the dynamics at short times. However, this has almost no effect on the final deposit as shown by \cite{ionescu:2015}, where they use the full Navier-Stokes equations and impose the movement of the gate as a moving wall boundary condition.  With the model proposed here, we also obtain this gate effect at initial time for small slopes ($\theta=0^\circ,9.78^\circ$), whereas the results are almost identical for large slopes ($\theta=16^\circ,19^\circ,22^\circ$).

An important result is the fact that our non-hydrostatic model predicts the parabolic shape of the velocity of the front as a function of time, as observed in the experiments. This is not reproduced at all by hydrostatic models where the velocity of the front starts from its maximum. Such improvement is obtained even though our model is only weakly non-hydrostatic, in the sense that we do not take into account the contribution of the stress tensor in the non-hydrostatic pressure.

\section*{Acknowledgements}
This research has been partially supported by the Spanish Government and FEDER through the research projects MTM2015-70490-C2-2-R and RTI2018-096064-B-C22, and by the ERC contract ERC-CG-2013-PE10-617472 SLIDEQUAKES. The authors would like to thank Cipriano Escalante for the interesting discussions related to this work.

\bibliographystyle{plain}

\bibliography{Biblio}
\end{document}